\documentclass[nonacm,sigconf,9pt]{acmart}

\usepackage{graphicx}
\usepackage{todonotes}
\usepackage{acronym}
\usepackage{url}
\usepackage{xspace}
\usepackage{mdframed}
\usepackage{subcaption}
\usepackage{lmodern}

\newcommand{\magic}{Revelio\xspace}

\acrodef{TVM}[TworthyVM]{trustworthy confidential virtual machine}
\acrodef{VM}[VM]{virtual machine}
\acrodef{SME}[SME]{Secure Memory Encryption}
\acrodef{SEV}[SEV]{Secure Encrypted Virtualization}
\acrodef{SNP}[SEV-SNP]{SEV Secure Nested Paging}
\acrodef{ES}[SEV-ES]{SEV Encrypted State}
\acrodef{SP}[AMD-SP]{AMD Secure Processor}
\acrodef{SGX}[SGX]{Software Guard Extensions}
\acrodef{TDX}[TDX]{Trust Domain Extensions}
\acrodef{CCA}[CCA]{Confidential Compute Architecture}
\acrodef{TEE}[TEE]{trusted execution environment}
\acrodef{IC}[IC]{Internet Computer}
\acrodef{TCB}[TCB]{trusted computing base}
\acrodef{VCEK}[VCEK]{Versioned Chip Endorsement Key}
\acrodef{ASK}[ASK]{AMD SEV Key}
\acrodef{ARK}[ARK]{AMD Root Key}
\acrodef{TLS}[TLS]{Transport Layer Security}
\acrodef{SSL}[SSL]{Secure Socket Layer}
\acrodef{CA}[CA]{Certificate Authority}
\acrodef{HRoT}[HRoT]{Hardware Root of Trust}

\usepackage{caption}
\begin{document}
\copyrightyear{2023}
\acmYear{2023}
\setcopyright{acmlicensed}
\acmConference[Middleware '23]{24th International Middleware Conference}{December  11--15, 2023}{Bologna, Italy}
\acmBooktitle{24th International Middleware Conference (Middleware '23), December  11--15, 2023, Bologna, Italy}
\acmPrice{15.00}
\acmDOI{10.1145/3590140.3629124}
\acmISBN{979-8-4007-0177-1/23/12}
\title{Trustworthy confidential virtual machines for the masses}

\author{Anna Galanou}
\authornote{Research was conducted while working at DFINITY Foundation}
\orcid{0002-4148-7631}
\affiliation{%
  \institution{TU Dresden}
  \country{Germany}
}
  \email{anna.galanou@tu-dresden.de}

\author{Khushboo Bindlish}
\orcid{0009-0004-5718-1532}
\affiliation{%
  \institution{DFINITY Foundation}
  \country{Switzerland}
}
  \email{khushboo.bindlish@dfinity.org}

\author{Luca Preibsch}
\orcid{0009-0005-9755-3502}
\affiliation{%
  \institution{Friedrich-Alexander-Universit{\"a}t Erlangen-N{\"u}rnberg}
  \country{Germany}
}
  \email{luca.preibsch@fau.de}
  
\author{Yvonne-Anne Pignolet}
\orcid{0000-0003-0837-7948}
\affiliation{%
  \institution{DFINITY Foundation}
  \country{Switzerland}
}
  \email{yvonneanne@dfinity.org}

\author{Christof Fetzer}
\orcid{0000-0001-8240-5420}
\affiliation{%
 \institution{TU Dresden}
  \country{Germany}
}
  \email{christof.fetzer@tu-dresden.de}

\author{R{\"u}diger Kapitza}
\orcid{0000-0002-8116-7763}
\affiliation{%
  \institution{DFINITY Foundation}
  \country{Switzerland}
}
\affiliation{%
  \institution{Friedrich-Alexander-Universit{\"a}t Erlangen-N{\"u}rnberg}
  \country{Germany}
}
  \email{ruediger.kapitza@fau.de}



%!TEX root = ./main.tex

\begin{abstract}
Confidential computing alleviates the concerns of distrustful customers by removing the cloud provider from their \acl{TCB} and resolves their disincentive to migrate their workloads to the cloud. 
This is facilitated by new hardware extensions, like AMD's \ac{SNP}, which can run a whole \acl{VM} with confidentiality and integrity protection against a potentially malicious hypervisor owned by an untrusted cloud provider.
However, the assurance of such protection to either the service providers deploying sensitive workloads or the end-users passing sensitive data to services requires sending proof to the interested parties. 
Service providers can retrieve such proof by performing remote attestation 
%and verifying that their workload runs on authentic hardware and has been initialized as expected.
while end-users have typically no means to acquire this proof or validate its correctness and therefore have to rely on the trustworthiness of the service providers.

In this paper, we present \emph{\magic}, an approach that features two main contributions: i) it allows confidential \ac{VM}-based workloads to be designed and deployed in a way that disallows any tampering even by the service providers and ii) it empowers users to easily validate their integrity.
In particular, we focus on web-facing workloads, protect them leveraging \ac{SNP}, and enable end-users to remotely attest them seamlessly each time a new web session is established.
%We validate \emph{\magic} by utilizing \ac{SNP}, present in the 3rd-generation of AMD's EPYC CPUs, to secure the \acp{VM} and by providing a lean, yet powerful, web extension.
%
To highlight the benefits of \emph{\magic}, we discuss how a standalone stateful \ac{VM} that hosts an open-source collaboration office suite can be secured and present a replicated protocol proxy that enables commodity users to securely access the Internet Computer, a decentralized blockchain infrastructure.
\end{abstract}
\maketitle
%!TEX root = ./main.tex

\section{Introduction}

% Mission: Motivation behind confidential computing 
%Since the advent of cloud computing, customers face the dilemma of whether or not to trust the cloud with their sensitive code and data.
% 
%Responding to such reservations, cloud providers are currently rolling out \emph{confidential computing}~\cite{CC21} leveraging  novel security extensions offered by multiple hardware vendors which aim to reduce the \ac{TCB} that customers inherently have to accept when they migrate their workloads. 
%By protecting these workloads within hardware-isolated compartments, called \emph{\acp{TEE}}, customers are ensured that their data remain shielded from any unauthorised access of privileged system software layers, such as the hypervisor, and the cloud-managing personnel. 

Responding to cloud customers' concerns over the security of their workloads, cloud providers are currently rolling out offerings with novel security extensions enabled, facilitating the use of hardware-isolated compartments, called \emph{\acp{TEE}}.
By migrating their security-sensitive workloads inside TEEs, customers are ensured that their data remain shielded from any unauthorised access of privileged system software layers, such as the hypervisor, and the cloud-managing personnel. 
In recent years, VM-based TEEs, where entire \acp{VM} can be isolated, have gained a lot of traction from the industry because of their seamless deployability, leveraging technologies like AMD's \ac{SNP}~\cite{snp-white}, Intel's \ac{TDX}~\cite{TDX} and ARM's \ac{CCA}~\cite{arm-cca}.

% Mission: VMs are the future. 
%One of the first widely available technologies that enabled confidential computing was Intel’s \ac{SGX}, in which the segregation in secure and unsecure parts happens at the process level. 
%Even though assuming an untrusted OS reduces the attack surface, it also necessitates the use of specialized tools that either offer semi-automated partition~\cite{glamdring17atc} or a seamless deployment of entire legacy applications inside containers~\cite{scone16OSDI, graphene17sgx}, both of which can be cumbersome for users and are linked to several other issues, like performance penalties.
%To overcome these challenges and close the compatibility gap with the existing IT landscape, major hardware vendors nowadays focus on protecting an entire \ac{VM} and offer \ac{VM}-based \acp{TEE}, such as AMD's \ac{SNP}~\cite{snp-white}, Intel's \ac{TDX}~\cite{TDX} and ARM's \ac{CCA}~\cite{arm-cca}. 
%Such an approach is increasingly favored and may soon prevail among the confidential computing offerings as it enables customers to deploy a \ac{TEE} rather seamlessly, rendering it primarily a matter of \ac{VM} image customization.

% Msg: Make clear that the potential of confidential VMs is not fully exploited. E.g., end users are currently out of scope. 
Although confidential computing is considered a big step forward in terms of cloud security, its benefits cannot be reaped to a maximum degree without additional effort. 
In particular, end-users of \ac{VM}-based TEEs are unaware if a service they access is secured via trusted execution or served by a commodity environment. 
% Next, we expose escalating the pain points: 
% Problem 1: No common API!
They could be assured of these guarantees via remote attestation, which essentially provides proof for the authenticity of the underlying hardware as well as for the integrity of the software loaded inside the \ac{TEE}.
However, to our knowledge, there is currently no standardised, widely established approach to provide remote attestation as a service to end-users besides the option to expose directly the hardware manufacturer-provided APIs that are typically accessible only to the \ac{VM} owner/service provider. 
% Problem 2: What is the sense of a hash we don't understand? 

When it comes to verifying the integrity of the \ac{TEE} state, the end-users need to be aware of what constitutes an acceptable state rather than just depending on a \ac{CA} to validate the \ac{TEE} authenticity.
The \ac{TEE} state is reflected on the attestation report, which includes a cryptographic hash over the \ac{VM}'s initial memory context right before its launch. 
This hash is usually compared against a pre-computed expected value and upon a successful match, the validating party can be reassured that the \ac{VM} has been initialised in a known good state.  
This is sufficient for service owners who intend to verify that a \ac{VM} runs within a \ac{TEE} as expected and they have full control and knowledge over the provided \ac{VM}'s image.

However, for end-users the situation is different and they are largely in the dark regarding the implementation and configuration of the services they access. 
%This is also the case when the service is based on open-source software.
%
% Problem 3: Even if we would know the VM is under the control of the service provider 
Besides that, even if they had access to the \ac{VM}'s source code and therefore be in the position to validate the hash of the \ac{VM}'s initial state against an expected value, they would inherently have to trust the service provider not to tamper with the \ac{VM} after it has been measured, via management APIs (e.g., ssh) or directly boot it with a malicious kernel for instance. 
The latter scenario is a real threat since the \ac{VM}'s initial state typically only includes the virtual firmware used to boot the machine and nothing more. % and the VM is later on customized.  
Therefore, the measurement included in the attestation report is not indicative of either the operating system or the root filesystem's state that the \ac{VM} booted with.
Consequently, by solely having access to a remote attestation report, which contains an initial \ac{VM} context, end-users cannot really verify if their data is securely handled.

%\rk{I don't get the next paragraph...if we have a client-side sw that we can validate -- end-to-end does not require TEE I think. Thus, the next paragraph is tricky.}
The implicit trust in the service provider cannot be avoided even if end-to-end encryption is in place, as is in the case of email services (e.g. ProtonMail), collaboration suites (e.g. CryptPad), cloud storage services (e.g. Tresorit), search engines (e.g. DuckDuckGo) etc. 
Even though this technique provides a high level of protection for user data and their privacy, such guarantees are ensured as long as the endpoints are not compromised.
If the server-side software, for instance, has vulnerabilities, an attacker can manage to execute code remotely and retrieve login credentials, or any other sensitive information. 
Therefore, verifying the state of such applications and enlightening the end-users about them is of significant value.

%% Nice but not necessary: 
%Ignoring the aforementioned issues, up to now the use of trusted execution secured \ac{VM} remains largely hidden. 
%In the best case, an end user could assume that, for example, based on an HTTPs connection to a website, the service provider has performed remote attestation and injected %the necessary SSL certificates to give users access to the service \cite{brenner19serverless}. 
%However, this is only a vague indicator.

%\begin{figure}[t]
%\begin{center}
%  \includegraphics[scale=0.42]{figures/motivation.png}
%\caption{\magic's main idea: In case the end-users intend to access a web-based service that is running on a confidential \ac{VM}, there is no way that they can verify the security guarantees promised by the service provider. That means that the service may not be running on a confidential \ac{VM} at all, or it might be insecure due to missing security patches. Also, the data is under the full control of the service provider.   
%\magic enables to validate the state of the \ac{VM} via the convenient way of a web extension so that the end-users can verify the utilized service.}
%\label{motivation}
%\end{center}
%\end{figure}

% High level goal: 
In this paper, we present \emph{\magic}; %an approach tailored to end-users facilitating deep validation of confidential \acp{VM} and their offered services.
an approach that enables end-users to easily, systematically, and deterministically validate a \ac{VM} and its hosted services, so that they no longer need to entrust service providers with the integrity and confidentiality of their data. % (Fig.\ref{motivation}).
Service providers manage the \ac{VM} and its services, but are averted from either tampering with an attested \ac{VM} or manipulating it during booting. 
In particular, we describe how to reduce their privileges to a level, where they can only perform denial of service to the \ac{VM} but cannot access users' data or tamper with it.  
The end-users on the other hand are enabled to retrieve a complete fingerprint of the \ac{VM}'s initial state and attest it. 
Since only a small fraction of the users will have the technical skills or the time to determine if a \ac{VM} and its offered services are properly implemented and securely configured, we discuss how the assessment can be delegated to a third party.
This allows less technically savvy users to benefit from knowledgeable users or institutions that can perform the validation of the system on their behalf.
We detail the steps on how this can be achieved, especially for web-based services, and show how commodity users who are entirely unaware of trusted execution implicitly profit from our system.

\emph{\magic}'s contributions can be summarised as follows:

\begin{itemize}
	\item We present an approach that enables seamless and automated remote attestation when accessing a \emph{\magic \ac{VM}}-hosted web-facing service via a web extension. 
	Depending on the user and the service provider, the validation can be performed by a third party, such as an auditing company, a community using a blockchain infrastructure, or the end-users themselves. 
	
	\item We introduce and implement a systematic way to disallow service providers from accessing or even altering users' data.
	To achieve that, we securely configure the \ac{VM} network-wise, render its root filesystem, containing the services of interest, immutable and extend its initial cryptographic fingerprint to cover the necessary state so that end-users receive a representative report of the attested \ac{VM}.  
	
	\item We highlight how \magic can address a wide variety of use cases, as well as elaborate on %the implementation of a state-less replicated \ac{VM} that can be horizontally scaled.
%	The latter resembles a protocol translation proxy that enables access to the Internet Computer, a decentralised blockchain infrastructure. 
how it can protect a stateful open-source collaboration office suite and a protocol translation proxy that enables access to the Internet Computer, a decentralised blockchain infrastructure.
\end{itemize} 
% This is also mentioned later, should we remove it?
We implemented \magic leveraging AMD's \ac{SNP}~\cite{snp-white}, however, upcoming \ac{VM}-based TEEs, such as TDX~\cite{TDX} and ARM's \ac{CCA}~\cite{arm-cca} can also be alternatives for our approach.
Therefore, \magic can be deployed in a hardware-agnostic fashion, as long the TEE follows the \ac{VM} model. 
%
%{ \color{red}
%The remainder of the paper is structured as follows: 
%
%Section~\ref{sec:bg} highlights the essentials of trusted execution required to implement \magic by the example of AMD \ac{SNP}.
%Section~\ref{sec:scenarios} discusses targeted application scenarios.
%Next, Section~\ref{sec:design} ...
%}   

%!TEX root = ./main.tex

\section{Background}
\label{sec:bg}

This section provides background information on the AMD \ac{SNP} technology, a boot method of the confidential \acp{VM}, as well as the basic workflow of \ac{SSL} certificate generation for web servers.
While the first two build the essential basis for \magic, the latter part is needed to be adapted so that the control of service providers over the \magic \acp{VM} can be reduced.

\subsection{AMD \ac{SEV} Technology}
%AMD trusted execution technology for \acp{VM}~\cite{sev-white} is now available in the third generation. 
\magic is relying on AMD \ac{SNP} technology to guarantee the confidentiality and integrity of \ac{VM}'s memory and hence of the services running on it.
%It started as a x86 technology and feature extension of AMD's \ac{SME}, targeting to isolate virtualized environments from the hypervisor.
% 
%In contrast to \ac{SME}, 
\ac{SEV} uses a unique memory encryption key for each \ac{VM} and tags it with an address space identifier, preventing cross-\ac{TEE} attacks and unauthorised usage inside the processor. 
The encryption keys are generated by a firmware running on a dedicated processor called, \ac{SP}. 
The encryption is transparent to the hypervisor, performed inside dedicated hardware in the on-die memory controller that encrypts data when it is written to DRAM and decrypts it when read.
% 
%In order to prevent information leakage through the general purpose registers, which were not encrypted upon \texttt{VMEXITS}~\cite{sev-secan} with \ac{ES}~\cite{sev-es-white} all CPU register contents get encrypted when a \ac{VM} stops running. 
%
%\acf{SNP}~\cite{snp-white} is the successor to the \ac{ES} iteration that adds memory integrity protection preventing hypervisor-based attacks like data replay, memory re-mapping and addressing rollback attacks on the remote attestation protocol~\cite{sev-attack-ra}. 
%We leverage \ac{SNP} for \magic because it fulfills our requirement for encrypted and integrity protected \ac{VM} memory.

\subsubsection{Remote Attestation on \ac{SNP}}
% This is for the attestation and at the ready for request property
Besides the \ac{VM} protection, \ac{SNP} provides proof to the user about the correct \ac{VM} deployment and the authenticity of the \ac{SEV} hardware by performing remote attestation.
The provided proof is called an attestation report and holds several pieces of information, among which is the VM's measurement; a cryptographic hash of the components that form the initial \ac{VM}'s state and is being taken by the trusted firmware running on \ac{SP}.
It also contains the unique identifier of the processor, called Chip ID, as well as the TCB version, which refers to the version number of the SNP firmware.
To allow the authentication of an \ac{SEV} platform and enable the remote attestation protocol, \ac{SEV} signs the attestation report with the 
%chip endorsement key (CEK), which is unique to the local platform and fused in the die of the processor. 
%
%\rk{Next sentence is unclear.}
%For \ac{SNP} th is derived chip encryption key from the version of the \ac{SP} firmware signs the cryptographic hash and ensures the guest owner that it came from a certain \ac{VM}.
\ac{VCEK}, which is unique to the local platform's processor and takes into consideration the TCB version.
While the data contained in the attestation report are fixed during VM's lifetime, there is a protected path between the \ac{SP} and the \ac{VM} so that arbitrary data can be added to the \ac{SP}-signed attestation report via the field \texttt{REPORT\_DATA} and be cryptographically linked to it.

\subsubsection{Measured direct boot}
\label{sec:meas-boot}
\begin{figure}[t]
\begin{center}
  \includegraphics[width=\linewidth]{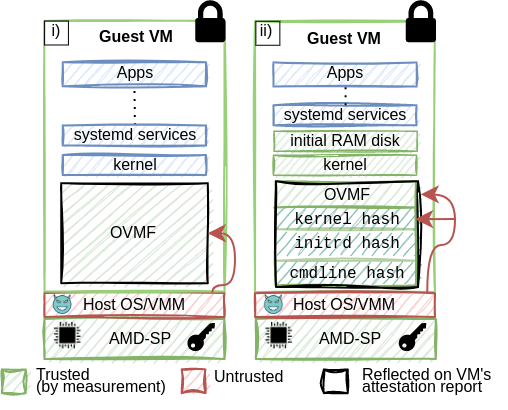}
\caption{Direct boot (i) vs measured direct boot (ii).}
\label{Meas-dir-boot}
\end{center}
%\vspace{-2mm}
\end{figure} 
%Despite the key management, secure internal communication and attestation mechanism that \ac{SNP} offers, the provided support targets an owner approved boot process.  
%
After the \ac{VM} has finished loading, \ac{SP} finalises its hash measurement,  %reflecting the current initial state.
% 
%The \ac{VM} owner can attest the measurement at this point and if it matches an expected value, they can inject a key into the \ac{VM}'s context that will be subsequently used to decrypt an attached volume and boot the \ac{VM}. 
%While this approach is rather generic and easy to implement it gives the \ac{VM} owner, and hence the service provider in case of \magic 's model, far too much control over the \ac{VM} as they can arbitrarily decide on the booted system configuration. 
%  
%Another significant issue is that the measurement contained in the attestation report typically represents only the very first code that is loaded during the first stage of \ac{VM}'s launch. 
%In case that QEMU is used, that is equivalent to the first QEMU firmware volume and nothing else. 
which typically corresponds only to the first virtual firmware volume and nothing else.
To implement \magic, an approach is required that 
%enables an independent boot-process of a \ac{VM} and more importantly one that 
imprints the exact initial state including any persistent state, used to bring up the \ac{VM}, in the attestation report.
This is achieved by an approach called \emph{measured direct boot}~\cite{sev-measured-boot}, implemented as a series of patches to the hypervisor, QEMU, and to the virtual firmware, OVMF.
%Support for measured direct boot for SEV(-ES) is already upstream in OVMF and QEMU. 
The modified OVMF package creates space in the firmware binary (Fig.~\ref{Meas-dir-boot}) so that it can store a table with the hashes of the kernel, initial RAM disk (initrd) and the kernel command line. 
When QEMU boots a \ac{VM}, it hashes each of these blobs and injects the hashes into the special table. 
These components are passed by the hypervisor as arguments via the \texttt{fw\_config} interface. 
OVMF measures each component upon reception and compares the measurement to the ones placed into the designated table. 
In case any of the measurements do not match, the boot fails. 
The hypervisor is untrusted, but the firmware is measured and reflected on the attestation report, hence the hashes injected by the hypervisor can be verified by anyone performing remote attestation and having access to all sources of the \ac{VM}.

\subsubsection{Securely saving persistent data}
% secure encryption of persistent state
Besides remote attestation, \ac{SNP} provides the possibility to generate additional key material for data sealing %and a process of encrypting data in persistent memory to the \ac{VM}.
that can be derived from different sources depending on their use case.
In the most basic case, a \ac{VM} can request a sealing key derived from VM's measurement, which essentially makes it accessible only by a VM with an identical cryptographic fingerprint. 
The sealing key is exchanged via a trusted path between the \ac{SP} and the \ac{VM} and stays protected from any component outside of the TCB.

%
%Using such a sealing key, data can be encrypted and securely written to persistent memory. 
%In case of a shutdown, the sealing key together with all the rest of in-memory state will be lost.
%However, the \ac{VM} can retrieve the same key via the \ac{SP} and use it to unseal the data, since it is bound to its measurement.
%Together, this enables to transfer data securely across restarts of a \ac{VM} with the exact same initial state.
% 
%In order to transfer data across changes between different version of a \ac{VM}, a more flexible approach is needed. 
%One alternative is to tie the sealing key to an Identity Block, a guest owner-signed blob that contains fields that allows them to uniquely identify the \ac{VM} and is included as part of the attestation report.
%
%As we will see later, this flexibility is not without downsides, if the service providers should be strongly limited in their control over the internal state of a \ac{VM}. 
%
%Despite the flexible and far reaching mechanism for sealing offered by \ac{SNP}, similar to other trusted execution technologies such as \ac{SGX}, without further measures so call fork and rollback attacks are possible. 
%In essence, in context of a VM restart or upgrade the sealed disk state can be reset to a previous intact but outdated state. 
%Such attacks can be detected, for example, by involving users of the \ac{VM} \cite{lcm17dsn} or more directly prevented by additional witness machines \cite{rote17CSS}.

\subsection{Certificate generation and Certificate Authorities}
% Write about domain-validation certificates, Let's Encrypt and ACME clients, like certbot.
% Mention CSR, that are signed with the private key and suffice for proof of possession
% Explain wild-card certificates
One of \magic's key contributions is establishing trust to web-facing services with which end-users interact. End-users typically have no evidence to assess the security guarantees services promise to offer. 
Nowadays, the vast majority of these services use the hypertext transfer protocol secure (HTTPS); a secure version of HTTP, the primary protocol used to send data between a web browser and a website. 
%
% We could try to make this shorter ...
HTTPS uses an encryption protocol to encrypt communications, called \ac{TLS}, the successor protocol of \acf{SSL}. 
Coupled with a Public Key Infrastructure (PKI), SSL/TLS provides verifiable identities via certificate chains and private communication. 
\magic leverages this to provide a binding between the web interface of the service and its cryptographic fingerprint imprinted on the attestation report.

%The private counterpart lives on the web server's premises, while the public key/certificate is available to anyone that wants to interact with the server in a secure fashion. 
When a browser connects to a SSL-secured site it will first retrieve its SSL certificate and check that it has not expired, it has been issued by a trusted \ac{CA} and that it is being used by the intended website. 
%If any one of these checks fail, the browser will display a warning to the end-user letting them know that the site is not secured by SSL. 
An SSL certificate is essentially a signed attestation binding a subject to a public key and can be issued by \acp{CA}, who in turn have their own certificates terminating at a small set of self-signed root certificates.
In order to obtain and install a certificate, a server operator is used to manually prove control of the domain name and complete a payment transaction to a \ac{CA}.  
Nowadays, \acp{CA} like Let's Encrypt~\cite{letsencrypt}, offer domain-validated certificates through a standard protocol at no cost to server operators. 

Let’s Encrypt is the first browser-trusted CA that established a standard protocol, called Automatic Certificate Management Environment (ACME), to automate identity validation and certificate issuance via clients, like certbot~\cite{certbot}, without intervention from web server operators or \ac{CA} staff. 
To obtain a certificate for the domain, the user can pass to the client a Certificate Signing Request (CSR)~\cite{csr} and ask Let’s Encrypt \ac{CA} to issue a certificate with a specified public key. 
The CSR includes a signature by the private key corresponding to the public key in the CSR, among other information like the domain, organisation, country name etc.
After the Let’s Encrypt CA has received the request, it verifies the signature and upon successful validation, it issues a certificate for the requested domain with the public key from the CSR and returns it to the client. 

%!TEX root = ./main.tex

\section{\magic's Design}
\label{sec:design}
In this section, we give more details about the scope of our work, the motivation behind it, as well as the threat model we adopt and the entities involved in it. Besides that, we are going to describe \magic's design by outlining our key requirements and elaborating more on the components of its architecture.

\subsection{Objective}
%The inherent risks involved in trusting the cloud with an enterprise's infrastructure, including insider threats and accidental misconfiguration of assets which can lead to exposure of enterprise data and annual damage of millions~\cite{cybercrime}, motivated the technology stakeholders to set up a new line of defense introducing confidential computing. 
%This essentially facilitates the use of hardware-isolated runtime environments that allow the execution of sensitive applications in a protected space and ensure the confidentiality as well as integrity of their related data-in-use. 
%With confidential computing offerings becoming ubiquitous, security-sensitive domains, like the eHealth space~\cite{ehealth} in some European countries, are already making using it a regulatory requirement. % and therefore we have deemed necessary for our system to meet with such demands.
%
\acp{TEE} provide hardware isolation that can either happen on a process-level boundary or on a VM-level one. 
On the former one, the operating system is treated as malicious which means that any syscall interrupt coming from the isolated code will immediately exit the \ac{TEE} context.
Consequently, that means that all the applications need retrofitting to work inside the \acp{TEE} which hinders the widespread adoption.
To that end, the industry is gradually moving towards VM-based \acp{TEE}, such as \ac{SNP}~\cite{snp-white}, 
%, Intel's \ac{TDX}~\cite{TDX} and ARM's \ac{CCA}~\cite{arm-cca}
where the kernel is part of the isolated code/VM and therefore the applications can seamlessly work without any modifications. 

With confidential VMs making inroads to the public cloud, we also opt for it as \magic's foundation, so that we can render our system easily deployable by real-world web-facing services.
For these kinds of services, the protection of their data-in-use leveraging hardware-based trusted execution is being decided by the service providers and typically communicated to the end-users to alleviate their security concerns.
However, the end-users themselves are largely unaware of the environment such provided services are running in and if the security measures that have been promised are actually in place. 
By virtue of this inherent lack of control, they have to implicitly place their trust in the service provider for the installed software and the utilised cloud infrastructure without receiving any evidence for it.

Assuming that such evidence, i.e. \ac{TEE} attestation report, was available to the end-users, they would have to be in a position to verify that an authentic Hardware Root of Trust (HRoT) generated it and that the integrity state of the utilised service matches the expected one.
The latter requires the evidence to contain information that reflects the service's state and from the users to have knowledge about what to accept as "correct" and "incorrect" state.

Considering that the whole foundation of the confidential computing paradigm lies in verifying before establishing trust, it is of utmost significance that the end-users have the power to do so.
However, currently, they can only take the security assurances from the service providers at face value and have no proof for those.
In our work, we address this problem of lack of proof as well as the challenges that come with exposing it to the end-users. More specifically, \magic
\begin{enumerate}
\item exposes attestation evidence to the end-users in a practical manner,
\item imprints the cryptographic fingerprint of the utilised service to this evidence,
\item applies measures to preserve the integrity of the service during runtime,
\item and enables end-users to verify the provided evidence.
\end{enumerate}

\subsection{Threat model}
Before describing how \magic tackles these challenges and fulfils our goals, it's necessary to elaborate more on the threat model we assume during each phase of the VM's lifetime, namely the provisioning, occupancy and decommissioning phase.
The main stakeholders %(Fig.\ref{cloud-hosting}) 
involved in those are the:
\begin{itemize}
	\item Cloud providers; they manage the cloud infrastructure in terms of hardware and the associated software (host OS, hypervisor, firmware), offering \acp{VM} to customers as Infrastructure as a Service (IaaS).
	\item Service providers; they take the role of a cloud customer renting \acp{VM} to provide services to end-users.
	They are in control of the VM's virtual firmware, OS and software.
	\item End-users; they use the services, as they please, and are largely unaware of technical details of the services' runtime system, like if it is cloud-based or what kind of hardware and software is being used. 
	They use a web browser with \magic's web extension installed.
\end{itemize}

%\begin{figure}[t]
%\begin{center}
%    \includegraphics[width=\columnwidth]{figures/actors.png}
%\caption{Main stakeholders in a cloud-based hosting scenario.}
%\label{cloud-hosting}
%\end{center}
%\end{figure}

\paragraph{Provisioning phase}
Before the confidential VM's deployment, its image needs to be built, configured and customised by the service provider so that it has the necessary software packages installed, as well as the network settings and security measures applied.
However, the end-users %do not trust that the VM has been configured with the correct services or any hardening measures to block intruders from entering it and 
assume that the \ac{VM} image may contain vulnerabilities to introduce backdoors or other malicious code planted by the service provider so that they can leak their data.
Besides the VM image itself, the host platform owned by the cloud provider has to be provisioned as well, so that it has the necessary firmware, kernel and hypervisor to support the \ac{TEE}.
The only component that is considered trusted on the host platform by the service provider and the end-users is the CPU hardware along with the \acl{SP} implementation.
The virtual firmware, kernel, initrd are assumed to be provided by the service provider to the cloud provider so that they can deploy the confidential VM later with the TEE-enlightened hypervisor.
Neither of those are considered trusted by the end-user since they can contain bugs either planted by the cloud provider to enable a privilege escalation from the host VMM or the service provider with the intention of retrieving sensitive data from the service.
%The service provider does not trust that the cloud provider has launched the VM with the correct kernel, initial RAM disk or virtual firmware either.

\paragraph{Occupancy phase}
In line with the confidential computing threat model, we assume that the cloud provider or a host platform intruder has full control over everything except the \ac{TEE} itself.
That basically includes the physical hardware as well as the entire software stack on the host platform, i.e. the OS and the hypervisor, which allows the adversary to perform a wide range of actions like modification of any file on the system,  %installation or removal of software 
and monitoring of system activity including network traffic, system logs, user activity etc.
We also consider man-in-the-middle attacks where an attacker can intercept network traffic between two parties, spoof or corrupt it, %inject malware or other malicious code into the intercepted traffic 
and redirect traffic to a different destination.
The cloud provider may also migrate the VM to a different physical host without the knowledge of the service provider exposing it in this way to security risks.
% Msg.: We go beyond the typical attacker model
%Besides this, we assume that a developer of code used inside a secure \ac{VM} could be malicious. 
%For example, she could try to introduce a back door in one or another form to the code. 
%Additionally, the \ac{VM} and service's managing personnel can be ill-intentioned and attempt to misconfigure the \ac{VM} in order to get full access to it and extract or manipulate user-specific data. 
% Msg.: Denial-of-service is out of scope
In our threat model, we do not target denial-of-service attacks, so a cloud provider's attempt to starve the VM of resources, such as CPU or memory, in an effort to disrupt its operation is out of scope. 
%
% Msg.: We need to say more about side channel attacks. 
Ciphertext side-channel attacks on the encrypted VM (by building a dictionary of plaintext-ciphertext pairs)~\cite{cipherleaks} are out of scope. 
%Side-channel attacks, like timing and page faults, are out of scope for the presented approach. 
%These are difficult to exploit in practice, and existing mitigation strategies introduce a high performance overhead and represent an independent line of research.
%The ultimate goal of \magic is to facilitate a systematic way for the end-users to verify the integrity of full software stack they access when using a service as well as the authenticity of the hardware below it. 
%The resulting trust relations can be briefly summarized as follows:
% Trust assumptions related to the cloud provider and the cloud operators 
%The cloud operators trust the utilized software and hardware. 
%This trust is to a large extent established by a security department, which validates the utilized software and hardware stack and reacts to newly identified vulnerabilities and attacks.
The cloud providers explicitly do not trust the service providers, as they may attempt to get unrestricted access to the cloud resources or tamper with the code and data of other tenants.
The end-users assume that the service provider may launch a malicious service in order to retrieve their sensitive data, or/and modify the environment to serve this purpose.
%\magic aims to change this by rendering confidential computing visible and beneficial to end-users, and thereby enabling the extensive verification of the utilized service. 
%In other words, the users can get a complete picture of how the service they use is implemented and configured. 
%Therefore, the no longer have to trust the service provider or the cloud operator. 
%Besides that, the service provider can be relieved from the inherent responsibilities of having direct and unrestricted access in the users' data.

%\paragraph{Maintenance phase}
% Remote access, debuging utilities, ssh access, software updates
%This may involve applying software patches and updates, troubleshooting issues, and providing support to the service provider by the cloud provider
%Cloud providers may offer maintenance and updates for the underlying operating system running on the VM, which can include security patches, bug fixes, and other updates to ensure the VM is secure and up-to-date.
%Such cases require a reboot of the VM after which the service providers or the end-users do not trust that the VM is in the expected state.
%We also assume that any remote access tools used for maintenance may be vulnerable to exploitation, potentially allowing attackers to gain unauthorized access to the system.

\paragraph{Decommissioning phase}
Ensuring the confidentiality of tenants and their secrets also after the node release is essential, so we address any attacks that can be launched by subsequent software running on the node aiming to retrieve any remaining information from the persistent storage.

\subsection{Requirements for \magic}
%In order to enable an end-user to extensively validate a \ac{VM}, we need to establish a set of requirements, pertaining to the functionality of the \magic approach, as well as its deployability in multiple use-case scenarios.
Our main objective is to remove the cloud provider from the trust domain of the end-users that use confidential web-facing services as well as reveal if those have potentially been misconfigured by the service providers, narrowing the trust computing base even further. 
On that account, \magic offers end-users cryptographic proof of the state of the utilised service as well as of the confidential VM it is running on top of.
To achieve that goal, our system needs to fulfil the following requirements pertaining to functionality (F1-6) as well as deployability (D1-3).

\textbf{F1: Evidence disclosure}
Before sending any sensitive information to the service, the end-users need to be able to receive cryptographic evidence about the service's state so they can attest it and assess its trustworthiness.
%\textbf{F1: Remote attestation} Before accessing the service provided by \magic, the user needs to be able to perform remote attestation of the \ac{VM}. 
%As an outcome of this process, an attestation report is generated which is essentially a hardware generated and signed proof containing the cryptographic hash of the VM's initial state. 
%\todo{ Why remove this? --> This includes the volatile and also non-volatile memory (such as any VM attached disk at startup).}

\textbf{F2: Trust only by measurement for TCB}
The evidence offered to the end-users needs to reflect the VM's volatile as well as non-volatile memory (i.e. any block storage device mounted at startup) so that they can have a representative view of the whole TEE that was loaded and not implicitly place their trust on its TCB.
This evidence should be cryptographically bound to the \ac{HRoT}.

\textbf{F3: Confidential VM \& service linkage}
End-users need to make sure that the evidence they receive corresponds to the confidential VM the service is actually running on.
Therefore, the confidential service's identity should be cryptographically linked to the hardware-based identity of the VM.

\textbf{F4: Integrity during runtime}
Aiming to avert any misconfiguration of the service by a malicious intruder or malevolent service provider, \magic should render it infeasible for them to modify the service as well as the runtime environment after booting, namely the network settings, ssh connectivity, approved ports etc.

\textbf{F5: Reproducibility and deterministic verifiability} 
Build verifiability is an important safety property for software releases \cite{reprod-report}, as external attesters can audit and verify if software is susceptible to various security problems introduced during the build process (e.g., surveillance malware, compromised cryptographic signatures, supply chain attacks, and untrusted dependencies).
By providing a reproducibly built image for the \ac{VM}, we intend to automatically accommodate a practical and efficient attestation by the interested parties as well.

\textbf{F6: Persistent state protection}
Considering that the service's persistent state may contain sensitive users' data, \magic should aim to protect the storage devices used by the confidential VM against any offline attacks performed by anyone outside of the TEE's context.

%\textbf{F7: Maintainability}
%Service's persistent state needs to be recoverable after any upgrades on the TEE software stack.

%\textbf{D1: Identification} End-users must be enabled to easily identify a \magic, when they access a new service. 

\textbf{D1: Ease of use}
\magic needs to enable end-users to perform remote attestation seamlessly as an integral part of the service utilisation. 
An intuitive way of presenting the attestation results and indicating possible security violations should also be provided by the system.

\textbf{D2: Flexibility over the verification}
By using \magic, end-users should be enabled to perform an extensive verification of the received evidence. 
However, considering the complexity of this task and the time investment it typically requires even for the technically-skilled users, a flexible approach needs to be provided so that they can either delegate the validation to third parties or perform it themselves.

\textbf{D3: Scalability}
Taking into account that \magic's use case scenario concerns web-facing services that need to be available at all times and serve a significant load of users' requests, \magic should take measures to support scalability.
	
\subsection{\magic in a nutshell}
To address the requirements set for our system, \magic has to be involved in the building/provisioning as well as deployment phase of the confidential service which takes place at the service and cloud provider's premises respectively.
In the following sections, we present the core parts of our design as well as the requirements each one of them aims to satisfy.

\subsubsection{Reproducible build as a basis for practical and efficient remote attestation}
%Taking into account that our main objective is to enable users to verify the state of the utilized service as well as of the platform it is running on, it is inarguable that the cryptographic measurement reflecting that state is of utmost importance. 
%Before deploying their VM on the cloud providers' infrastructure, service providers prepare the VM image during the building phase. 
%By adopting reproducibility in the build process, \magic averts the installation of backdoor-introducing malware on the service providers' VMs and detects any corrupted or outdated build environments that would have an impact on the expected measurements. 
Typically, in dynamic systems the system's state frequently changes due to system-centric events such as software package installations, dependencies in libraries, configuration of the build environments, timestamps, file permissions etc. 
This means that there may be multiple acceptable hashes for the \ac{VM} depending on the building environment, which makes the reconstruction of the expected hash and consequently the \ac{VM}'s attestation particularly cumbersome in the best case, and in the worst case completely unreliable. 
By removing any non-determinism in the build process, \magic guarantees identical built binaries and images for every invocation of the build process, so that the user can reproduce the expected measurement in a systematic and practical way (F5).%, thus satisfying requirement F5. 

\subsubsection{Robust configuration to ensure system's integrity during runtime}
Considering that we do not assume any runtime monitoring system running on the confidential VM, there is no way to ensure that the VM will stay intact after booting.
%Since in our system the run-time state is not reflected on the attestation report by virtue of not having any run-time monitoring system in place, we have to disallow any modification of the system's state that may happen either during the launch-time configuration of the VM or later during the normal operation. 
% Add some of the challenges/requirements that runtime monitoring systems have
% Cite keylime etc.
One of \magic's requirements (F4) is to take the necessary measures to guarantee integrity during runtime.
% To ensure that goal, we impose the following measures during the build process which can then be attested by the end-users.
We address this problem in two ways; first by blocking any inward connections to the confidential VM and second by enforcing a read-only root filesystem and rendering it integrity protected.
The first ensures us that no outside attacker or authorised personnel can access the \ac{VM} after booting and therefore they cannot spawn a malicious process or tamper with the existing ones during runtime.
% Cite similar attacks
The second one guarantees the integrity of the volume containing the root filesystem (including kernel, services etc.) against corruption by malware or persistent rootkits that hold onto root privileges and compromise devices.
%Since during launch-time the VM is being configured depending on the service provider's requirements, we have to ensure that ill-intended configuration cannot have any impact on the integrity of the VM and in worst case it can only lead to Denial-of-Service attacks. 
%This is achieved by sanitizing the configuration data that are being fed at the VM during the first launch.
%For the integrity protection, we perform a process known as verified boot leveraging dm-verity. 
%This is a block-device integrity solution that supports read-only integrity verification in kernel. 
%Instead of performing full file system verification in advance, it can be done on demand from a verified kernel, whose measurement we attest during the direct measured boot. 
%A transparent block device will be layered between the run-time system and all running processes. 
%Each block that is accessed via the transparent block device layer will be checked against a cryptographic hash which is stored in a central collection of hashes (hash-tree).
%The root of this hash-tree will be imprinted on the kernel command line arguments and consequently it will be reflected on the hashes contained in the firmware's hash-table. 
Both of those measures are being applied during build time so that the generated VM image has already the correct network configuration and the system integrity is protected.

\subsubsection{Expanding coverage of the reported state}
Based on the current state of the upstream versions of the hypervisor and virtual firmware that support the confidential VM loading, only the initial state of the VM is subject to the measuring performed by the trusted hardware, which essentially contains only the firmware. 
To expand what is covered by the reported state in the attestation evidence sent to the users (F2), we deploy the method of measured direct boot (~\ref{sec:meas-boot}). 
This allows us to imprint in the VM's final measurement, not only the virtual firmware's hash but the ones corresponding to kernel, initrd and kernel's command line respectively.
To extend the trust chain even further so that it includes the cryptographic state of the root filesystem, we pass as kernel command line argument the root hash of the block device's integrity metadata.
This has already been generated during build time, when the image of the rootfs is being constructed.
The code enforcing the integrity protection for the rootfs is part of the initrd and the kernel, which are both measured by the hardware in our system.

\subsubsection{Enabling end-users to attest before use}
% Argument - why end-users need to attest -> sensitive data to service, TLS: locks out mitm but server side can still be malicious
%		   - why we choose web-extension as the way to do it
% Disentagle single point of trust from CAs
% Need to serve the evidence (F1) but also make it convenient (D1)
% Report can be fetched and compared against pre-determined value(next subsection)
% Convenience; a) end-users need to know before they pass their data
%			   b) Integrate it to browser interface
%\magic tackles the problem of end-users being unaware of the state of the web-facing services they are using by facilitating a seamless delivery of attestation evidence to them before they pass sensitive data (e.g. account credentials etc.) to a potentially untrusted service.
%In order to provide a seamless attestation of the platform that the service is running, we provide a tool implemented as web-extension that requests an attestation report from the VM and validates it against an expected measurement.
%The main goal that drives our design is making attestation evidence available to the end-users, so that they can assess themselves if the service they are using is trustworthy or not.
End-users should be able to attest the service before interacting with it and passing their sensitive data, like account credentials, financial details etc.
Even if TLS is in place and assumed to provide authentication, integrity and privacy for the data transmitted across the untrusted channel, those are no longer guaranteed if the server endpoint is subverted by a potentially malicious service provider.
Therefore, by receiving actual proof that the server is in a good state end-users can benefit from the minimised TCB guaranteed by the confidential computing paradigm.
To serve both of the requirements F1 and D1, \magic offers the attestation evidence via a web extension, so that end-users can perform the attestation using the same web browser interface.

\subsubsection{Binding web-service's identity to the TEE}
An SSL certificate is essentially the digital identity of the web-facing service end-users interact with.
To verify this identity they rely on the \acp{CA} that have provided the root and intermediate certificates.
This root of trust is independent of the \ac{HRoT} corresponding to the attestation evidence and therefore the end-users cannot originate the received attestation evidence from the confidential VM that generated them.
\magic addresses this problem (F3) by cryptographically binding the TLS identity of the service to the TEE that it is running on, thereby enabling the end-users to verify that the proof they receive about the service is not only authentic but represents the actual service they interact with.

\subsubsection{TLS key sharing to serve scalability}
% Describe the need to generate inside the encrypted VM the priv key and have the DNS credentials elsewhere.
% Also describe the private key creation as a requirement?
% We also need to talk about private key sharing protocol --mutual attestation
%\todo{We need first to explain why the key-pair needs to be created inside the \ac{VM} next we can escalate to the topic on many \acp{VM}}
Since our system is meant to be deployed in decentralised secure services for which scalability (D3) is crucial, rate limiting on the SSL certificate creation is a critical problem that we need to take into account.
More specifically, certificate authorities like Let's Encrypt, consider a rate limit \cite{rate-limits} on the SSL certificates generated in a certain period to ensure fair usage among the users.
To address this problem, we assume that all the \magic-\acp{VM} that host instance of a service share the same SSL certificate.
The account credentials for the certificate generation reside in a platform belonging to the service provider's infrastructure and they are isolated from the public cloud.
This machine will be the one performing the DNS challenges for proving ownership of the service domain and distribute the generated certificate among the nodes after attesting them.

The SSL certificate will be generated based on a key pair of one of the nodes picked by the service provider's machine. 
This key pair essentially will be the TLS identity corresponding to the service that end-users are going to interact with later on and will be created in the context of one of the confidential VMs the service is running on.
The private counterpart of this pair will be later distributed among the rest of the nodes, so that they can serve the user's requests in a secure connection.
Before that, %the node owning the key pair corresponding to the SSL certificate 
 the TLS key owner will attest them in a mutual attestation protocol (Section~\ref{sec:implementation}) and encrypt its private key with the public one of the other attested node's unique pair.

\subsubsection{Tailoring verification of \magic VMs}
%\todo{Maybe we should more provide a continuum of choices with this outline cases as an extreme case.}
% The first way is not demanding and can accommodate users that do not have a strong technical background or have a more relaxed trust model. 
% The "golden" cryptographic hash can be uploaded on a public repository from the service-provider and retrieved from there by the web-extension. 
% The end-user may choose to trust this measurement, considering that it can be verified and reviewed by the open-source community. 
% \todo{Here we already carve in stone that the kernel is provided as binary -- if we send to a security venue this is dangerous -- we should write it more generic.}
% The alternative requires from the enduser to retrieve the binaries installed on the initial VM image, namely the firmware, kernel, and initrd, be enlightened about the VM's configuration (i.e CPU model, number of cores, kernel command-line arguments) and reconstruct the measurement by using another provided tool. 
% Regarding the root hash of the rootfs' integrity metadata, it would be necessary to build the image of the rootfs, generate them from scratch and include the root hash as part of the kernel cmd given as input in the above-mentioned tool. 
% The latter one could be also avoided and the root hash can be retrieved in a similar manner, as the one explained on the first way, from a public repository. 
In order to make \magic more deployable in the context of web-facing confidential services that are being used by a wide range of end-users, we have designed it so that the attestation process can be tailored to fit users' trust boundaries (D2).
More specifically, after \magic delivers cryptographic proof to them about the state of the service, they have to compare it against the expected one(s).
If they have some technical expertise, they can reconstruct the state of the service on their own premises fetching the corresponding open-source components, compute the final cryptographic measurement and therefore make their own deduction on what state is "good" or not.
In case end-users cannot follow this process due to a lack of knowledge for instance, then they can retrieve the "golden" values from a third-party source.
This can come either from an auditing company that has been delegated with the task to check the software stack of the confidential VM for bugs and vulnerabilities, or from an on-chain decentralised autonomous organisation  (e.g. Internet Computer's Network Nervous System~\cite{ic}) where the community votes on the "good" values of the software running on the respective TEEs.

%\subsubsection{System support for upgrades}
%\ag{Is this supposed to be an integral part of our system? Or is it just a nice additional feature?}
% Begin by explaining the need for upgrades
% Reboots are not possible --qemu-system-x86_64: cpus are not resettable, terminating
% Since a reboot via reset is done by resetting the vCPU registers from the VMM, the VMRUN will fail since the guest registers are altered outside of the guest.
% The upgrade has to be done necessarily with a poweroff, and we need to store persistently the keys for the encrypted/sealed disks.
% Describe the scenarios where each sealing policy is more fitting.
% Begin with the measurement-based key derivation and proceed with some of the rest.

\subsubsection{Protecting persistent state}
To satisfy requirement F6 and protect any sensitive persistent data between shutdowns of the confidential VM, \magic leverages the feature of sealing and encrypts the external volumes with the VM's measurement-derived key.
In this way, no other VM that has not been in the expected state can decrypt the disk and retrieve any sensitive data belonging to the service, like the private TLS key.

\section{Use cases of \magic}
While arbitrary services can be protected by trusted execution, the use of \magic imposes an additional requirement for transparency. 
At a high level, a service provider has to make all code and relevant operational configuration data of their built image available to external parties. 
For VM-based \acp{TEE} this goes beyond the actual service code and includes the execution environment itself, like the OS.  
In the scenario where the end-users' trust boundaries are very narrow, it would be them who will retrieve and verify all of the components comprising the trusted execution context; in a less restricted setting an external authorised party can be given access and perform validation on their behalf.
%It has to be noted that the actual service data stays confidential at all times, due the use of trusted execution, and in accordance to the VM's imposed rules. 

Although providing access to all of the service's source code and operational configuration data to the public is not a common practice for commodity offerings up until recently, we argue that the shift towards the confidential computing paradigm might render it necessary. 
In particular, end-users can find \magic significantly useful in scenarios, where they share sensitive data, such as medical records~\cite{ehealth}, and in the context of private social exchange. 
There are also numerous other scenarios where the demand for the service's integrity might be of key interest, like in auction sites, lotteries and any form of e-commerce service.
%Also \magic facilitates to continuously validate data protection rules such as GRDP either by end-users or third parties.
When it comes to the execution environment of the services, like the OS, the firmware etc., the ever-growing trend of open source software~\cite{oss-eu,oss-enterprise,oss-cyber} already fulfils \magic's requirement for code availability.
%Nowadays OSS has become pervasive in data centers, consumer devices and enterprises allowing for greater transparency and accountability. 
%Since the source code is available to the public, anyone can review it to ensure that it is free of bugs, vulnerabilities, or malicious code which can later lead to faster patching and greater overall security. 
%First of all, if a service is already open source or depends to a large extent on open source code, we argue that proclaiming this information requires very little or not effort at all. 
%\todo{Actually I'm not sure what do you mean by that}
%Furthermore, from a validation perspective one can delegate the assessment of the trustworthiness of the code also to the providing community.

In this section, we discuss two use cases: 
\begin{itemize}
	\item The first example describes \magic's use to secure a cloud end-to-end encrypted collaboration suite, where data confidentiality is of key interest. 
	\item The second one refers to the need for confidentiality and integrity of the end-user-provided data, shows a greater level of complexity and represents a deployed instance of a protocol translation proxy. 
\end{itemize}

\subsection{End-to-end user-encrypted cloud collaboration suite with hardware-based trust}
Over the recent years
%, there are several applications that have integrated privacy enhancing technologies and cryptographic solutions to provide higher security standards for users in a seamless fashion. 
end-to-end encryption 
%in particular is one of the primitives that 
has significantly grown in demand as an effective means to prevent the compromise of an agent’s private information and is currently used by many messaging %applications, like Signal and Whatsapp.
protocols, like Matrix~\cite{matrix}, % is another example of a messaging protocol where end-to-end encryption is used, 
where users are ensured that messages can’t be spoofed and that only the senders and receivers of them can read the contents.
Recent research has shown though that a malicious homeserver can modify the execution environment~\cite{matrix-vuln}, i.e. deliberately add users to end-to-end encrypted rooms to decrypt future messages sent in the room or add a device in the room they wish to eavesdrop, both of which essentially invalidate the confidentiality guarantees promised by Matrix’ threat model.

Collaborative editing platforms, like CryptPad, also use end-to-end encryption and enable users to enforce access control to their data by themselves and not depend on a central authority for that, which usually coincides with the server hosting the content and is a single point of failure.
Despite the strong security guarantees that CryptPad offers, it still adopts a more relaxed threat model of an honest but curious cloud server~\cite{cryptpad} 
%and does not appear to guard against a server interfering with the document's list of participants or its history.
%In case that cryptpad is not hosted by the service provider but by the flagship instance instead, 
where one still must trust the validity of the Javascript hosted on it or sent to the end-users' browser~\cite{cryptpad-vuln, cryptpad-vuln2}, which can potentially leak sensitive information.
We argue that \magic can be used on that type of applications to fill this security gap and provide the users with the means to verify the trustworthiness of the software installed on the servers as well as preserve their data privacy and integrity.

\subsection{Protocol translation proxy: a use case for elevated security}
\label{BN}
\begin{figure}[t]
\begin{center}
  \includegraphics[width=\linewidth]{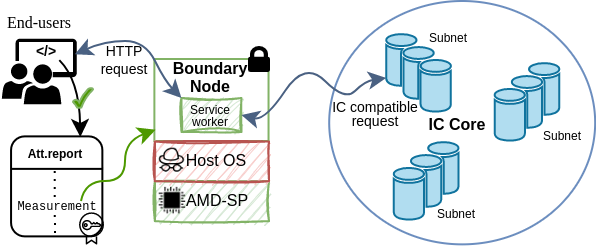}
\caption{\magic-protected Boundary Node.}
\label{bn}
\end{center}
%\vspace{-2mm}
\end{figure}
%\rk{This and the other figures look a bit blurry -- show pixel. Can you also produced revelio-bn.pdf from your tool and it would look better? If this is not possible I can live with it.}
% Explain the IC:
The \ac{IC}\footnote{\url{https://internetcomputer.org/}}~\cite{ic} is a decentralised platform for the execution of general-purpose decentralized applications in the form of so-called \emph{smart contracts}. A smart contract is a computer program that executes a certain program logic in a decentralized manner. Smart contracts on the \ac{IC} are called \emph{canisters}; 
all canisters are hosted on dedicated node machines running the \ac{IC} protocol. The protocol uses a threshold-signature-based agreement protocol to provide Byzantine fault tolerance. 
To achieve high scalability the nodes are partitioned in subnets.
To securely interact with a canister the \ac{IC} protocol has to be used, which enables the exchange of threshold-signed messages with the end-user.
As expected of a modern Web3 infrastructure, end-users typically access canisters delivering feature-rich web applications via a browser. 

%\rk{Should we say also a word about state? And updates?}

% Why is there a tranlation proxy? 
While the \ac{IC} provides the basis for digital democracy and Byzantine fault-tolerance, as well as fits into the conventional Web on a functional level, client-side software and browsers have yet to fully adopt the \ac{IC} protocol. 
The \ac{IC} overcomes this issue by utilising the Boundary Nodes that take the role of protocol translation proxies. 
A Boundary Node is capable of translating an ordinary HTTP request to an \ac{IC} protocol-compatible message exchange. 
This can happen either directly by simply receiving the HTTP request of an end-user, transforming it to an \ac{IC} message and forwarding it to the right canisters or as optionally via a JavaScript-based service worker. 
This service worker is returned by a Boundary Node on the first request of a user to the \ac{IC}.
Once activated on the browser side, ordinary requests are directly transformed to \ac{IC}  messages inside the browser by the service worker and forwarded to a Boundary Node which in turn delivers them to the \ac{IC}. 

% The BNs a case of \magic ;-)
The outlined approach gives any user immediate access to the IC via commodity browsers, however, it also represents a security risk as a malicious Boundary Node could manipulate user requests either directly inside the \ac{VM} or indirectly by providing a modified service worker.
The consequences of such a malicious node could be significant as it compromises the Byzantine fault-tolerance of the \ac{IC}. %and in the worst case, it could lead to a situation where end-users lose access to their accounts and digital assets.  
Letting Boundary Nodes run inside encrypted VMs facilitates higher security guarantees. 
By leveraging \magic, end-users can verify in a practical manner those guarantees, check if the services of the Boundary Node actually run in a hardware-based TEE and more importantly validate the exact software stack as well the configuration used on the node (Fig.~\ref{bn}). 
%Another great benefit of \magic, is that it locks service operators out of the \ac{VM} thereby averting insider attacks as elaborated in Section~\ref{}.

%!TEX root = ./main.tex
\section{Implementation details of \magic}
\label{sec:implementation}
%\magic is comprised by three components, namely the encrypted VM that runs on the cloud-provider platform, the attestation service running as part of the service-provider's workload inside the encrypted \ac{VM} and the web-extension that is enabled on the end-user's browser.
%
%In this section, we will provide implementation details for \magic's prototype pertaining to the components on the cloud provider's, service provider's and end-user's premises respectively.
In this section we will provide implementation details for the main components that \magic is comprised of, as described in Section~\ref{sec:design}, and we structure them by the phases of a \magic VM's lifetime, namely the provisioning of its image, its bootstrapping during the first deployment and its normal operation.
For our prototype, we leverage the \ac{SNP} hardware for the seamless encryption of the \acp{VM}, however, our approach can be based on any VM-based \ac{TEE}, including Intel's \ac{TDX} \cite{TDX} or ARM's Realms \cite{arm-cca}.

%\subsection{Cloud-provider's infrastructure}
%\subsubsection{Hardware}
%\rk{Needs to be shorter and more to the point.}
%In order to provide the highest level of protection for the service's workload inside the encrypted VM and comply with our threat model, which considers an untrusted host and hypervisor, we deploy a processor from the third generation of EPYC servers (Milan series) that has the AMD SEV SNP feature enabled. The cloud provider needs to have the relevant BIOS settings enabled so that the cloud tenants can leverage the feature for their VMs.
%\subsubsection{Kernel}
%\rk{At the moment I'm a bit unclear if this is implementation or evaluation? At the moment it feels more as these details fit better in the evaluation.}
%Besides the BIOS settings, the host platform needs to have a 5.19 or latest kernel version installed to facilitate the communication with the AMD Cryptographic Coprocessor for the SEV-SNP guest management, as well include the required boot arguments in the kernel command line.

%\subsubsection{Hypervisor and OVMF firmware}
%For our prototype we leverage qemu as the VMM, and this also need to have specific patches installed to support not only the SEV-SNP API but the direct measured boot as well. The latter one is implemented in the hypervisor and the firmware as well.
%\rk{This misses references to sources.}
%\subsection{Service-provider's workload}
\subsection{Image provisioning}
Before the VM's deployment in the cloud provider's infrastructure, the service provider needs to build and configure its image accordingly so that it complies with our requirements. This step lays the groundwork for the trust chain establishment and \magic's bootstrapping later on.
%\rk{This is a bit abstract and even less detailed than maybe the design section. Is there nothing of interest to be said about the implementation. For the moment it feels a bit redundant -- taking here about Ubuntu etc. is maybe also not helpful.}

\subsubsection{Reproducible build}
%\rk{We are quite generic here -- I guess we have to discuss what is going on here?}
The \magic VM's image is essentially an on-disk combination of an \ac{SNP}-aware Linux kernel binary, an initrd which contains a root filesystem with all the user-land programs that have been produced in the service provider's CI/CD pipeline as well as their corresponding software dependencies and libraries.
Since we assume that the VM is being launched with the method of direct boot, a bootloader is not necessary to exist in the image.
To achieve reproducibility, we deploy a deterministic build process which generates repeatedly equivalent images given the same set of source code files, build scripts, and build environment leveraging bazel~\cite{bazel} and its hermeticity.
Furthermore, the build scripts are modified in order to remediate sources of non-determinism (e.g., timestamps, build paths, file ordering and permissions) by clearing all files that may lead to in-deterministic build (e.g. /var/lib/apt/lists/*, /var/lib/dbus/machine-id/ etc.), squashing all timestamps and specifying a uuid for each partition we create.
Besides that, we create two docker images; the first one is meant to pull the needed packages and build the software dependencies while the second is meant to hold the actual binaries that we need which we copy from the former one. 
In that way, we don't actually pollute the final image with non-deterministic elements.

%
%\rk{The next two sentences miss a bit context - how do they relate to the implementation section.}
%To verify that the generated artifacts are identical across all build processes, in every build we compute their cryptographic hashes and compare them against the expected ones, that are stored in a trusted registry on the service provider's infrastructure.
%This verification is being performed regularly to test the reproducability of the live releases.

%\rk{I think you need to start from a different point. My idea would be the git repository and develop it from there.}
Besides the non-determinism sources related to the filesystem building, application packages can be another source of such problems since the package versions can change on every invocation of apt-get, dnf etc., thereby leading to different builds.
%Besides that, considering the fact that the APT packages can be updated in the upstream APT repository and therefore lead to different builds, during the build process we pull a published docker image instead of installing the packages from scratch.
To tackle this problem instead of installing the packages from scratch during every build, we pull a published image instead (i.e., docker image).
This image needs to be updated regularly and be publicly available so that it can be retrieved in every build.
We also need to ensure that it is trusted and integrity protected, so every time the software dependencies change, the new image is built in a protected environment (i.e. Gitlab runner) and then pushed to the registry.
%All the images are being built against the git revision that is published in a trusted registry by the service provider.
Finally, it should be noted that our system uses as a starting point an official Ubuntu 20.04 LTS Docker image, which is maintained by Canonical and therefore we trust the latter to a certain extent. 
However the sources and the build processes of such images are public, so they can be subjected to an audit and a review from the community.
\begin{figure}[t]
\begin{center}
  \includegraphics[width=\linewidth]{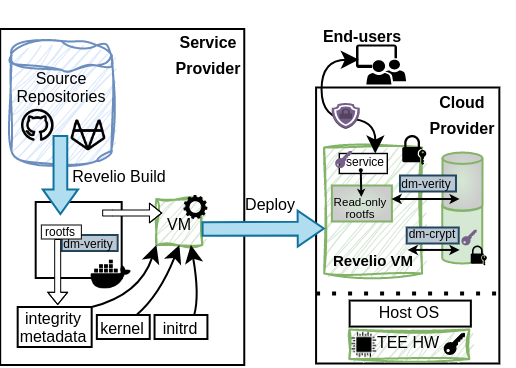}
\caption{\magic VM's build and deployment.}
\label{arch}
\end{center}
%\vspace{-2mm}
\end{figure} 

\subsubsection{Extending the measured envelope}
Leveraging direct measured boot\footnote{\url{https://www.mail-archive.com/qemu-devel@nongnu.org/msg945567.html}}~\cite{sev-measured-boot} allows us to extend the measured envelope from the virtual firmware up to the kernel, initrd and boot arguments. 
In order to extend this even further and establish a trust chain up to the system's rootfs, we include the root value of the rootfs' integrity metadata to the kernel's command line. 
The root value along with the rest of metadata of the underlying block device is generated during the build process (Fig.~\ref{arch}), when the rootfs is being constructed, leveraging the dm-verity utility of the Linux kernel.
Dm-verity~\cite{verity} essentially uses a Merkle tree of sha256 hashes computed for the device's blocks and verifies them every time they are being read, ensuring that files have not changed between reboots or during runtime.
%The hash tree for the rootfs is planted inside the disk image along with the root hash.
%The kernel and initrd binaries are generated as separate files during the building, since they are needed as arguments to the qemu for the direct measured boot.
When dm-verity is being setup for the hardware storage device on which rootfs has been stored, the latter is being remapped through a target driver to produce a new, transparent, virtual storage device.
We modify VM's init process so that it mounts the read-only and integrity-protected new virtual storage device corresponding to the rootfs, retrieving the root hash from the kernel command line and the rest of the integrity metadata from the designated partition of our choosing.
Since initrd is measured as well in our method, we ensure that if any of those measures are not being applied, it is going to be reflected on the attestation evidence of the~VM.

\subsubsection{Blocking unauthorized inward connections}
%\rk{I guess this needs to be a bit expanded. It is not ssh but one need to make security analysis for all services that heir interface cannot be used to arbitrary change the service. What about the config image. The latter is something we did not inherit from IBM.}
%The attestation performed on the VM takes into account only the load-time measurements taken by the AMD-SP during the first boot. Since our system does not rely on any integrity monitor during runtime, we need to ensure that there are not going to be any unapproved changes after the VM has been booted.
During the build phase, we configure the network services accordingly so that no unauthorised connections are allowed to come through the VM.
%On this end, we configure the VM to not accept any ssh connections from outside and hence avert any remote access from a potential malicious user.
The network configuration is part of the rootfs and the relevant services are contained in initrd, therefore their integrity is guaranteed with the dm-verity utility and the measurements will be reflected on the attestation report via the direct measured boot.
%\todo{Maybe we should skip this subsection, since no real implementation details are mentioned here?}
\subsection{Bootstrapping}
During the first boot of the \magic VM, there are certain services that are being triggered pertaining to the encryption and integrity protection of the disk as well as the TLS identity creation.

\subsubsection{Disk encryption and integrity protection}
%\rk{I was wondering (have not recently read the whole paper) if we need to explain that we are using sealing to support reboots. The latter is important for stateful services. We do not address the case of image upgrades ;-)}
Complying with our threat model necessitates the encryption of the secure-sensitive partitions of our disk and the integrity protection of the ones that do not contain confidential data. 
Regarding encryption, we leverage the sealing capability of the \ac{SNP} hardware and derive a key based on the VM's cryptographic measurement.
We choose that policy to ensure that only untampered VMs on the same platform can successfully decrypt the disks and access the data.
This is particularly useful if we need to persist data after a shutdown or because of an unplanned power outage.
During the first boot, this key is derived on the fly and encrypts the chosen volume (Fig.~\ref{arch}) leveraging Linux kernel's feature, dm-crypt, and cryptsetup~\cite{cryptsetup} to set it up. 
Regarding integrity protection, as previously mentioned, we modified initramfs to setup dm-verity for the rootfs and boot with the virtual device as the root partition instead of the ordinary one.
Before mounting, we leverage veritysetup~\cite{cryptsetup} to verify the integrity of the block device, passing the root hash from the kernel command line and the rest of the integrity metadata that have been planted on another partition during the build phase.
If the verification is successful we can proceed with the booting, otherwise, the VM's launching is terminated.

%\subsubsection{Persistent state handling}
%Since the VCPU registers are being encrypted when SEV-ES or SEV-SNP is leveraged, rebooting our VM is not possible so we need to power it down.
%In case we need to persist data after a power down, or because of an unplanned power outage, we leverage the sealing feature of SEV hardware.
%Similarly to the encryption of the partitions, we choose the measurement as a policy for the key derivation.
%However, in case of an upgrade the use of the measurement as a sealing policy is no longer a viable option.
%For that reason, the Family-ID can be used for which the service provider needs to create an ID-Block containing this information and sign it with a key owned exclusively by them.
%This has to be sent at the host so that it can be used along with the other arguments in the qemu command to launch the encrypted VM.

\subsubsection{Unique VM identity creation}
During the first boot, a service is triggered that creates a key pair unique to each VM.
This will either be the TLS identity that will correspond to the SSL certificate identifying the web-facing service, or it can be used for secure data exchange between VMs after a mutual attestation has taken place.
%The key pair is stored in one of the partitions that are encrypted with the sealing key derived from the VM's measurement so that it can be protected.
%This key pair will be either used to generate an SSL certificate or to encrypt the private key of the chosen node.
After the key pair creation, a set of attestation reports are being created.
The first one includes as \texttt{REPORT\_DATA} the hash of the public counterpart of the VM's identity.
The second report is related to the SSL certificate issuing, for which a Certificate Signing Request (CSR) is created first for the VM's key pair based on the configuration details on the service's domain.
The \texttt{REPORT\_DATA} contains the hash of the CSR in this case.
%Since all of the nodes will be attested by the secure machine and/or the node owning the SSL certificate private key, the service will also create a set of attestation reports containing the hash of the node's CSR and the hash of the node's public key respectively.
%The first one will be sent to the secure machine and the latter one to the chosen node as part of the mutual attestation.

\subsection{Normal operation}
%After the VM's image building in the service provider's infrastructure, the system is ready to be deployed in the cloud node so that the service can start and serve the end-users.
%\rk{As in the previous section -- what about configuration?}
After \magic's bootstrapping, the \acp{VM} on which the web-facing service will run, need to acquire the SSL certificate and the corresponding private key, since they all have to share them to satisfy our scalability requirement.
In our implementation, we assume that there is an isolated node running on the service provider's premises that holds the DNS API credentials and is responsible for the SSL certificate issuing, distribution and node attestation.
We will refer to this node as SP node in the following sections.
For the HTTP server we use nginx and to process the HTTP GET and POST requests as well as trigger the \magic relevant processes, we use CGI scripts that interact with the server via the FastCGI protocol~\citep{fastcgi}.
To validate the certificate chain while performing remote attestation, we contact AMD Key Distribution Server (KDS) \footnote{Hosted at \url{https://kdsintf.amd.com/}} and retrieve the root certificates (for \ac{ARK} and \ac{ASK}) as well as the VCEK certificate after we have specified the Chip ID and TCB version.
To validate the measurements of the reports, we compare them against some hard-coded values that have been planted on the VMs at the build time.
However, this can be changed and each node can contact a remote Trusted Registry that is maintained on a blockchain infrastructure for instance~\cite{ic}, where the community votes on what is a "good" state or not.

%\begin{figure*}
%	\centering
%	\begin{minipage}{0.5\textwidth}
%		\centering
%		\frame{\includegraphics[width=.95\columnwidth]{figures/ssl-cert.png}}
%		\caption{SSL certificate distribution.}
%		\label{SSL-cert}
%	\end{minipage}%
%	\begin{minipage}{0.5\textwidth}
%		\centering
%		\frame{\includegraphics[width=.95\columnwidth]{figures/mutual-att.png}}
%		\caption{Mutual attestation \& TLS private key distribution.}
%		\label{Mutual-att}
%	\end{minipage}
%\end{figure*}

%\begin{figure}[t]
%\begin{center}
%  \includegraphics[width=\linewidth]{figures/ssl-cert.png}
%\caption{Node attestation, SSL certificate generation and distribution.}
%\label{SSL-cert}
%\end{center}
%\end{figure}

\begin{figure}[t]
\begin{center}\hspace*{-0.6cm}    
  \includegraphics[scale=0.29]{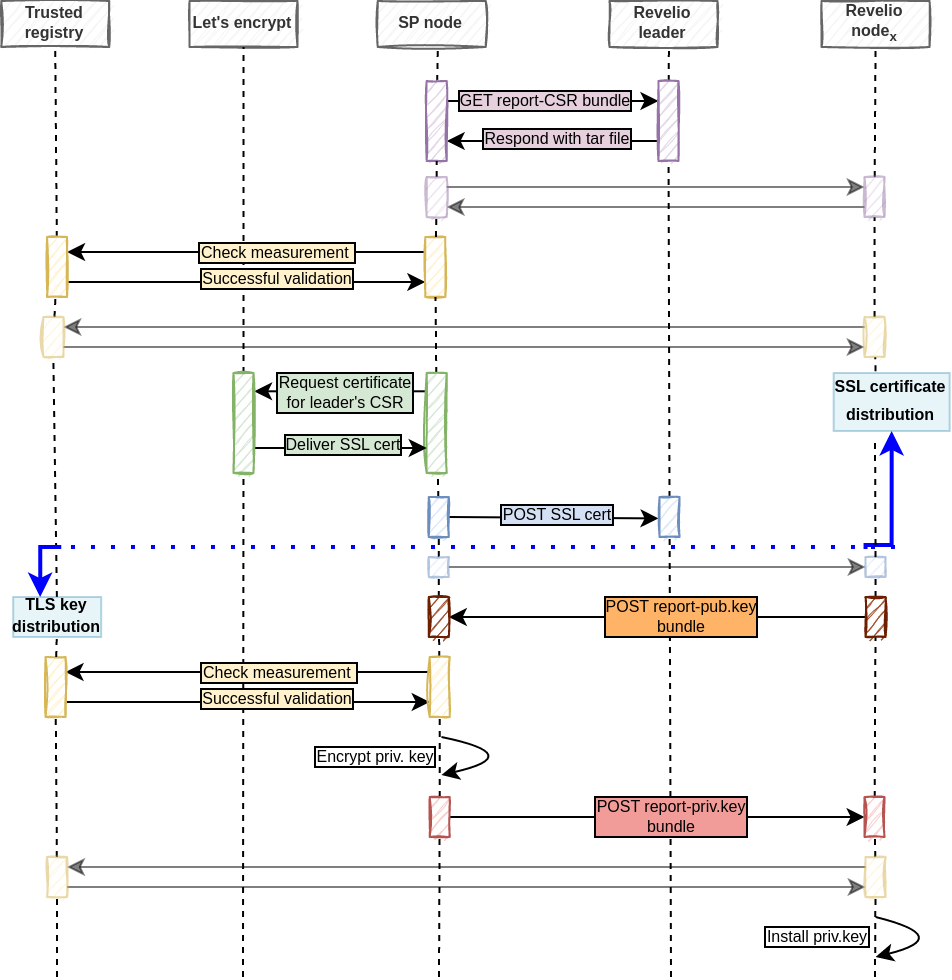}
\caption{SSL certificate and TLS private key distribution.}
\label{Mutual-att}
\end{center}
%\vspace{-3mm}
\end{figure} 

\subsubsection{Certificate management}
%The certificate management can be divided in three separate tasks, pertaining to the creation of unique key pairs for the nodes, the issuing as well distribution of the SSL certificate and the exchange of the private key corresponding to the certificate.
In order for the SP node to generate an SSL certificate, it needs to acquire first the CSRs from the nodes and pick one as the "leader".
Before entrusting one with this role, SP attests the whole set of the nodes by retrieving their report-CSR bundles (Fig.~\ref{Mutual-att}) and verifying two things; first that the measurements are as expected and second that the CSR's hash matches the one imprinted on the  \texttt{REPORT\_DATA} of the report.
%For the attestation, it performs HTTP GET requests and retrieves the reports, that contain the hash of the CSRs in the \texttt{REPORT\_DATA} field, along with the actual CSRs (Fig.\ref{Mutual-att}).
%To verify that the received CSRs belong to the nodes whose measurements are reflected into the attestation reports, the checksum of the former ones are compared against the  \texttt{REPORT\_DATA} of the reports.
%Besides that, the attestation of the nodes involves the validation of the node's IP, its certificate chain as well as of their measurement and Chip ID.
Besides that, during the attestation the node's Chip ID (imprinted in the report) and IP are being checked against a set of approved ones, to avert an impersonator from obtaining the private key of the SSL certificate, even if it presents an authentic and valid attestation report.
%The measurement validation provides the guarantee that the nodes have the correct setup and have not been maliciously tampered.
%Finally the certificate chain verification ensures that all the nodes have authentic AMD-SEV hardware.
After the round of attestations has been completed and the certificate has been generated for the leader's CSR, the SP node sends the certificate in a round of HTTP POST requests along with the leader's IP to notify the nodes, who they should contact later to acquire the private key.

For the secure exchange of private key (Fig.~\ref{Mutual-att}), each of the service provider's nodes contacts the chosen leader and performs an HTTP POST request with their report-public key bundle, containing a report with the hash of their public key, along with the public key itself.
The leader validates the report in a similar fashion as the SP node and then encrypts its private key with each node's public key.
Then it begins a round of POST requests sending its own attestation report bundle, containing the encrypted private key.
The nodes attest the leader in the same way and after verifying that the SSL certificate corresponds to the received private key, they install both of them in a temporary folder.
An incron job~\cite{incron} is triggered after that, installs the private key in the server-designated folder and restarts the HTTP server which is subsequently ready to serve requests.
It is worth noting that the private key is stored in an encrypted partition, so it cannot be leaked at rest.
%For the SSL certificate needed for the web-extension, we use let's encrypt \cite{letsencrypt} as our CA and certbot \cite{certbot} as the ACME client.

\subsubsection{Remote attestation by end-users}
To enable seamless remote attestation of a \magic \ac{VM} for end-users, the former must be integrated into the browser. 
This can be achieved in two ways, either by extending the browser itself or by a less intrusive way of a web-extension.
The former approach would require from major browser vendors to provide the necessary support for remote attestation. % at the current stage. 
For this reason, we designed a web extension that will automatically perform remote attestation for registered sites and enable it to learn about new sites offering \magic attestation while browsing the web.
Therefore, we assume that the validated HTTP server provides an attestation report under a well-known URL (e.g., as in the case of robots.txt~\cite{robots}) and runs in a \magic \ac{VM}.

\paragraph{Register \magic-conformed websites}
We assume two basic approaches of how a website and the associated \magic \acp{VM} can be registered via the \magic web extension. 
The first approach requires the end-user to manually register a website that should be validated via a configuration dialogue of the web extension.
In the basic case, the end-user needs to provide the domain name of the website and the expected measurement.
The measurement has either been computed by the user itself or it has been received via an out-of-band channel.
Here, more sophisticated schemes can be assumed but are out of the scope of this work. 
The second approach is simply to opportunistically learn about the \magic \acp{VM} while browsing the web. 
To do this, the web extension will detect if a website is hosted by a \magic \ac{VM} by checking the well-known URL for the attestation report. 
If an attestation report can be fetched, the web extension alerts the user who subsequently needs to validate the measurement.
The first approach is more secure and should be employed for security-sensitive sites since it does not require an initial untampered contact. 

\paragraph{Intercepting requests and performing remote attestation}
Assuming a domain has been registered with the web extension for remote attestation, the first access to it for a new browser context is always intercepted by the web extension and the attestation report is fetched from the \ac{VM}.  
% Check the HW ...
Next, the report is validated to verify that it originates from a \ac{SNP} secured context. 
After querying AMD KDS for the VCEK key, providing the corresponding Chip ID and TCB version extracted from the attestation report, 
%The AMD key server will respond providing the \acl{VCEK} key and the web-extension will validate %it has valid chain of signatures consisting of the \acl{ASK} and the \acl{ARK}. 
%As a next step it is validated that the attestation report is signed by the \ac{VCEK}. 
%This ensures that the context that provided the attestation report runs on genuine AMD \ac{SNP} hardware.
the web extension validates the certificate chain of VCEK (checking the ASK and ARK certificates) and that the report's signature matches VCEK's public key.
% Check remote context 
Following this, it validates the report's measurement against the locally stored one. 
If that succeeds, it is confirmed that the \magic \ac{VM} is considered trustworthy. 
% Check connection 
As a last step, it has to be validated that the browser has established a secure connection that terminates inside the \ac{VM}.
This is done via querying the browser regarding the public key of the current TLS connection used to connect to the remote site. 
If the value matches the public key that is part of the attestation report, it is ensured that the endpoint terminates securely in the \ac{VM}.  
Should any of the checks fail, this is flagged to the user and they have to make a decision to proceed with or abort the access to the website.
After this point, it has to be constantly monitored that the TLS connection is not reset and redirected to a different location. 
This is possible as the browser only validates if a connection request is established with a certificate that it trusts. 
A malicious service provider can, for example, create a new certificate as they control access to DNS and use this new certificate to redirect users away from the secure \ac{VM}.  
The web extension prevents this by intercepting all requests to the domain and validates that for each request the connection is based on the public key that was part of the attestation report. 

At a technical level, it has to be noted that currently only Mozilla Firefox provides the necessary APIs to run the web extension. The use of the service worker API (Section \ref{BN}) should be avoided for now since the (re-)loading of it can only be partially controlled. In this case, a small extension of the browser-provided APIs could lift this restriction.

%!TEX root = ./main.tex

\section{Evaluation}
In this section, we present a security analysis of \magic to demonstrate its robustness against attacks assumed in our threat model. Next, we assess the overhead that some of our \magic-enforced techniques impose in VM's booting, runtime as well as in the interaction with the web-facing service.

\subsection{Security Analysis}
In order to ascertain the protection of \magic, we provide a security analysis addressing potential attacks that can be launched by any of the stakeholders assumed in our model.

\subsubsection{Loading a modified kernel or initrd}
Since the hypervisor used to launch the VMs is under the cloud provider's control in the host platform, it can be instructed to load a malicious guest kernel %with no \ac{SNP} support for instance and ignore the one sent by the service provider.
%Similarly, it can load 
or a modified initrd which does not enable the integrity protection for the root file system or edit the kernel command-line arguments to pass a different root hash.
These attacks can also be performed by the service provider and are averted due to the measured direct boot deployed in our system. 
If the host uses the wrong kernel, initrd, or command-line, the measurements constructed by the QEMU and then verified by OVMF will not match and the booting will not be successful.
If the host replaces the OVMF with a malicious version that does not verify the hashes, then this will be reflected on the measurements taken by the \ac{SP} and hence on the attestation report.
If the host fills the expected hashes in the table constructed by the QEMU, but passes the wrong kernel/initrd/command-line, then the OVMF will detect that when it measures the individual components and checks them against the stored measurements so the booting will not be successful either.

\subsubsection{Tampering with rootfs}
\label{sec:mod-rootfs}
If the cloud/service provider modifies the rootfs image (with a malicious service for instance), then the root hash included in the kernel command line arguments will not correspond to it, so the mounting will be unsuccessful since we verify at this stage its integrity.
If they try to modify the root hash %included in kernel's command line 
as well, then this will be measured and reflected on the attestation report, so the attestation of the \ac{VM} will fail.

%\subsubsection{Malicious service}
%An ill-intended service provider may launch a \magic VM with a malicious web service and get a legitimate SSL certificate, since they have access to the DNS credentials for the domain.
%Even though the validation of the certificate chain will succeed, since the VM runs in authentic AMD hardware, the verification of its measurement will fail.
%This is because the service logic is imprinted in the rootfs and the hash of this is included in the measurement we acquire via the AMD-SP.

\subsubsection{Modifying the system during runtime}
An external attacker may attempt to modify the system during runtime since any changes after boot will not be reflected on the attestation report.
In order to do that they would have to remotely access the system which is not possible with our imposed system configuration, or modify the rootfs (see Section~\ref{sec:mod-rootfs}).
%If the attacker has access to the host platform, then they may try to modify the relevant configuration at boot time.
%These are part of the rootfs and therefore their integrity is guaranteed because of the dm-verity utility and the measured direct boot.
Even if those measures fail, since dm-verity protects the disk at a binary level, even a single bit change anywhere in the disk will cause dm-verity to raise errors, including mounting the disk read-write since that will cause small changes in metadata that get written to disk.

\subsubsection{Rollback attacks on the VM image}
An ill-intended service provider may launch a VM with an obsolete software stack so that they can exploit an existing bug.
%In that case, the certificate chain validation would succeed since the VM has been deployed in authentic AMD hardware, and the IP as well Chip ID verification will be successful too.
The certificate chain, Chip ID and IP validation would succeed, but the verification of the VM's measurement will fail, since we assume that the obsolete cryptographic hashes are being revoked every time there is a newer image rollout to prevent rollback attacks.

\subsection{System performance}
In this section, we evaluate how much delay \magic's techniques introduce to the VM's booting during bootstrapping as well as during runtime.  
%Revelio in terms of latency, demonstrating the performance overhead that it can introduce on the service provider's side as well at the end-users' side. 
%The first factor to evaluate it is going to be the booting overhead, then the bootstrapping, namely the pre-attestation, launch-attestation as well as any certificate generation and distribution that is necessary for Revelio's main operation.
Our measurements were taken on a machine equipped with an AMD EPYC 7313 16-core processor and 112\,GB RAM.
%,  and two 10 GB Ethernet network interface cards connected to a Gigabit switched network. 
We run Ubuntu 20.04.6 with Linux kernel 5.19.0-rc6 with the \ac{SNP} patches installed on the host, and the VM is running Ubuntu 20.04.5, a Linux kernel 5.17.0-rc6 with the \ac{SNP} patches installed.

\subsection{Booting latency}
%In the VM, the kernel starts up in 6.41s, while the userspace system services run in 2 mins and 197 s.
The services that are relevant to \magic's implementation, which we evaluate are (see Table~\ref{table:overhead}):
\begin{itemize}
\item the encryption service, which retrieves the VM's sealing key derived from its measurement, and then encrypts the chosen volume with cryptsetup,
\item the device-mapper service, which manages the integrity protected and readonly virtual volume created with veritysetup,
\item the rootfs verification service, which verifies its underlying volume with the provided integrity metadata and root hash,
\item and the VM identity creation service, which creates the VM's key pair, CSR, and a pair of reports holding this data.
\end{itemize}
%    1.011s cryptsetup@var_crypt.service (setup-encryption.service + setup-conf-encryption.service)                    
%     611ms setup-encryption.service /boot/config
%     442ms setup-conf-encryption.service 
%     219ms dev-mapper-vroot.device                                 
%     123ms setup-bn-key.service                   
%      41ms var.mount                              
%      19ms systemd-cryptsetup@vda10\x2dcrypt.service                
%      10ms systemd-fsck@dev-mapper-var_crypt.service                  
%       9ms systemd-tmpfiles-setup.service                         
%       8ms systemd-fsck@dev-mapper-store\x2dnginx\x2d\x2dcache.service
%       8ms systemd-fsck@dev-mapper-conf_crypt.service                 
%       7ms systemd-fsck@dev-mapper-store\x2dvar\x2d\x2dlog.service      
%       3ms boot-config.mount                                          
%       1ms boot.mount

\begin{table}[t]
\caption{\magic imposed delays on first boot}
\centering
\begin{tabular}{|c|cc|cc|} 
 
 \hline
 %& \textbf{Latency (ms)} & \textbf{Overhead} (\%)\\
 & \multicolumn{2}{c|}{\textbf{Latency (ms)}} & \multicolumn{2}{c|}{\textbf{Overhead (\%)}}\\
 \hline
 \hline
 & BN & CP & BN & CP\\
\texttt{dm-crypt} setup & 611 & 481 & 2.76 & 4.94\\%
\hline
\texttt{dm-verity} setup & 219 & 194 & 0.97 & 1.94\\%
\hline
\texttt{dm-verity} verify & 4680 & 3340 & 25.94 & 48.61\\%
\hline
Identity creation & 123 & 132 & 0.54 & 1.31\\%
\hline
\end{tabular}
%\vspace{-4mm}
\label{table:overhead}
\end{table}
%Considering that the encryption of block-devices as well as the generation of the integrity metadata for the root-file-system introduces some overhead, we intent to clarify the booting delays that \magic introduces.
%Starting with the enablement of the encryption of block-devices we consider as a baseline the booting time that is observed when none of our techniques is applied.
% Mention size of partition
The encryption service latency is size dependent and the volume it was performed on had a size of 84\,MB which was sufficient for both of our use cases.
%\rk{Why again 84\,MB? I think this value is for the BN VMs okay but would it be also realistic for the cryptpad? Maybe you should say that we used the BN partition which has a size of 84 because it has minimal state? Even if this is the case I was wondering why 84 MB? Is it the var partition you are referring to than you could write this?}
The same applies to the verification of the dm-verity protected rootfs, which in our case is 4\,GB.
The encryption of the volume along with the VM's identity creation happens only on the first boot.
The performance overheads, presented in Table~\ref{table:overhead}, have been calculated against the total time it took for a \magic-protected Boundary Node (BN) and \magic-protected Cryptpad Server (CP) to boot and initialise all the system services, which were %2mins and 9.239s (129239ms) and 18.152s respectively.}
%18.152s and 5.638s respectively.
22.725\,s and 10.211\,s respectively.
The difference in the booting time stems from the fact that the Boundary Node has a lot more services that are starting up during boot in comparison with the Cryptpad Server that has only the server instance and the \magic system-related services.
%2mins and 9.239s (129239ms) and 10211ms respectively.}

\subsubsection{I/O request latency}
To further evaluate the overhead that dm-crypt and dm-verity introduce during runtime, we simulate read and write requests in an encrypted volume of 10\,GB size and read the files under the read-only integrity protected rootfs of 4\,GB respectively.
We have configured dm-crypt to use \texttt{aes-xts-plain64} for the cipher and \texttt{pbkdf2} with 1000 iterations for its derivation.
For dm-verity, the chosen hash algorithm is \texttt{sha256} with a data and hash block size of 4\,kB.
%The reads and writes are performed with a block size of 4kB and an increasing number of blocks spanning from 1 to 65536.
%For the dm-crypt experiments, we evaluate the reads and writes in the O\_DIRECT and standard mode (annotated as direct/indirect on the plots).
%O\_DIRECT bypasses the I/O management of the kernel (including the page cache ) and the data are transfered directly between the user buffer and the file on the underlying block device.
%dm-crypt encrypts write requests before sending them further down the stack to the actual block device and decrypts read requests before sending them up to the file system driver.
%Direct I/O is not supported on verity files so we only evaluate it with the buffered I/O.

\begin{figure}[t]
\begin{center}
  \includegraphics[width=\linewidth]{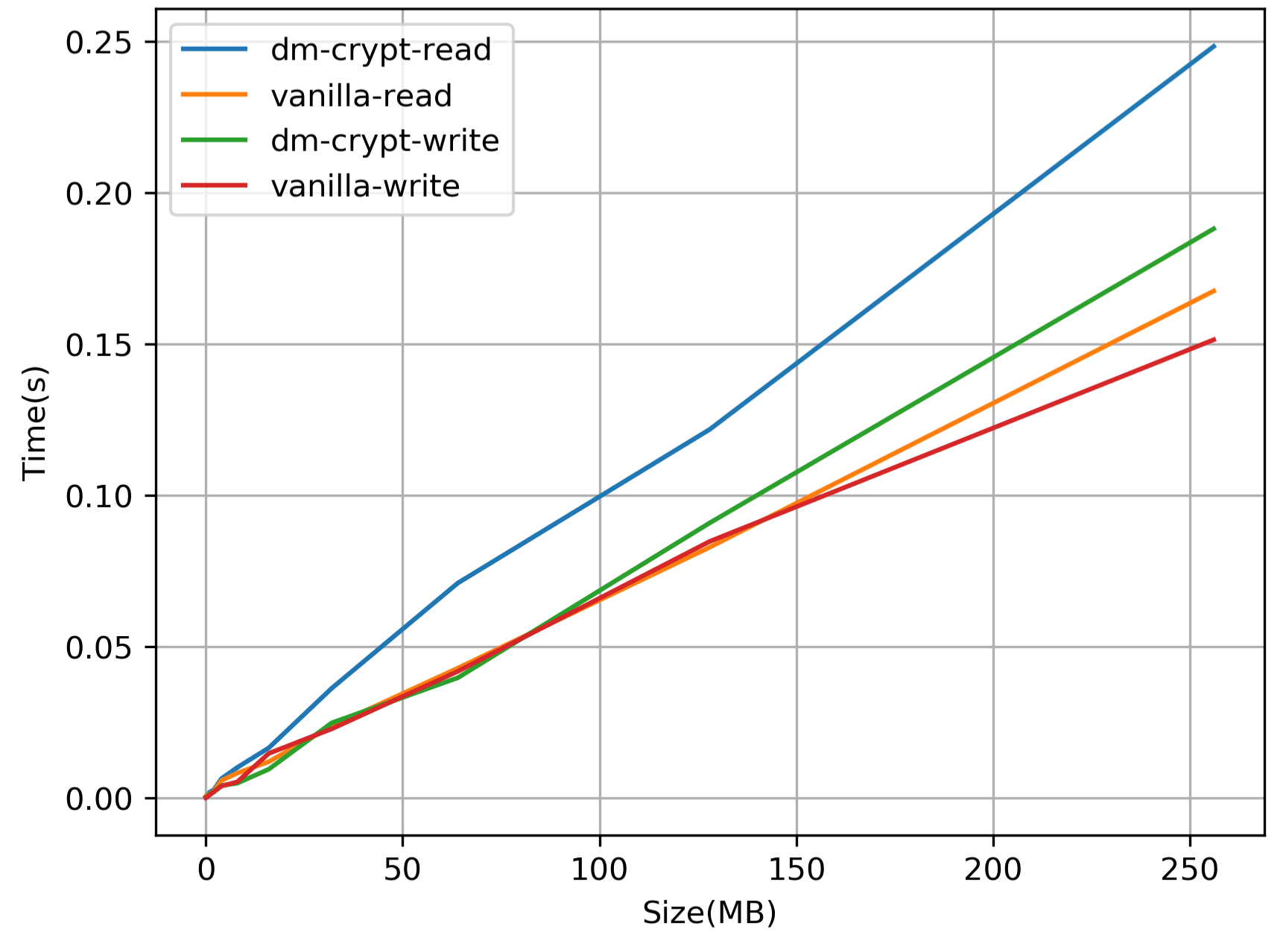}
\caption{Dm-crypt I/O latency.}
\label{dm-crypt-lat}
\end{center}
\end{figure}

\begin{figure}[t]
\begin{center}
  \includegraphics[width=\linewidth]{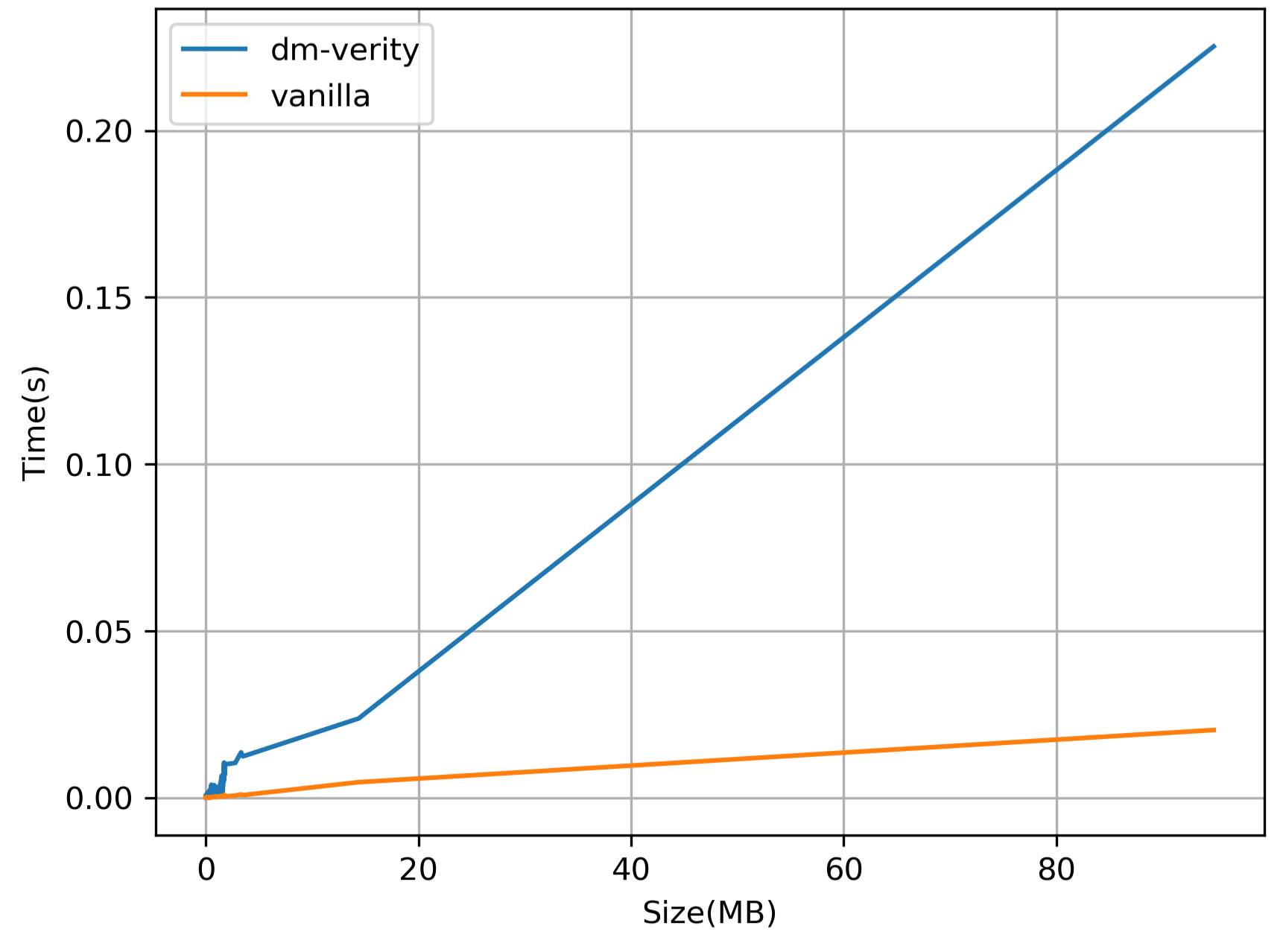}
\caption{Dm-verity read latency.}
\label{dm-verity-lat}
\end{center}
%\vspace{-2mm}
\end{figure}

The write and read requests are performed via the \texttt{dd} utility with a block size of 4\,kB and a total size of up to 256\,MB, because for both of our use cases, namely the Boundary Node  and the Cryptpad Server, the average file being read or written does not surpass this size. 
%Reading files of size up to 256mB from the encrypted volume doesn't seem to introduce any significant overhead (Fig.\ref{fig:dm-crypt-read}), with the minimum to be 16.39\% and the average 64.21\% in O\_DIRECT, while in the standard mode with buffered requests the minimum overhead drops to 1.99\% and the average to 26.32\%.
%For write requests, the minimum overhead is 0.77\% and the average 63.01\% in O\_DIRECT, while in the standard mode those are 0.35\% and 12.033\%.
%For bigger files up to 4gB, we notice a bigger overhead with the average to be rising to 32.72\% and 22.27\% for read and write requests respectively leveraging the I/O buffering which is the typical scenario.
For read requests the minimum overhead introduced by dm-crypt (Fig.~\ref{dm-crypt-lat}) is 1.99\% and the average is 26.32\%, while for write requests those are 0.35\% and 12.03\% respectively.
Considering the evaluation of the integrity-protected volume (Fig.~\ref{dm-verity-lat}), we read the files under the BN's rootfs where the biggest file has a size of 94.8\,MB 
%and the minimum and average overhead are 0.90\% and 835.38\% respectively.}
and the read latency presents on average a 9.35$\times$ slowdown.
%For files up to 4gB, the average rises to 97.11\%.}
%\rk{The figure say Time(ms) instead of Time~(ms) it should be the latter for both measurement figures. The same applies for Size(MB) --> Size (MB)}
%\rk{Why are we using sec if is always below a sec? This is unusual.  Why do we have a shorter figure description than the title? The title is redundant. I would use the title of the figure as the caption! Or even: Read measurement of dm-verity? }

%\rk{Can we have ms instead of s? And could we remove the title of the figure and make it the subtitle instead of  Dm-crypt latency. Also again Time(s) --> Time (s) or even better Time (ms) also Size(MB) --> Size (MB)}

%\begin{figure*}[htp]
%\centering
%\begin{subfigure}{.5\textwidth}
%  \centering
%  \includegraphics[width=\textwidth,height=0.55\linewidth]{figures/read_lat.png}
%  \caption{Read request latency}
%  \label{fig:dm-crypt-read}
%\end{subfigure}%
%\begin{subfigure}{.5\textwidth}
%  \centering
%  \includegraphics[width=\textwidth,height=0.55\linewidth]{figures/write_lat.png}
%  \caption{Write request latency}
%  \label{fig:dm-crypt-write}
%\end{subfigure}
%\caption{Dm-crypt latency on I/O requests}
%\label{fig:test}
%\end{figure*}
\subsubsection{SSL certificate operations latency}
%\rk{It does not say anything about the distribution of the machines -- can we have one more sentence here.}
Regarding the impact of the SSL certificate generation and distribution (see Table~\ref{table:ssl-lat}), we measured the time that takes the \ac{SP} to retrieve the attestation evidence from one node -- which comprises of the CSR reflecting its key and the attestation report containing the hash of the CSR--, validate the report's signature, certificate chain, measurement and hash, as well as the time to create the SSL certificate with certbot and distribute it to the node.
This happens typically once every 90 days when the SSL certificate needs to be renewed and redistributed, so it doesn't affect the runtime performance.
%\rk{If you have the tenths of a second for your SSL measurement Table~\ref{table:ssl-lat} pls add them for the table otherwise. It is what it is.}
\begin{table}[t]
\caption{SSL certificate generation and distribution}
\centering
\begin{tabular}{|l|r|} 
 
 \hline
  ~ & \textbf{Latency (ms)} \\
 \hline
 \hline
	{Attestation evidence retrieval} & 17  \\
\hline
   {Attestation evidence validation} & 13 \\
\hline
	{SSL certificate generation} & 2996 \\
\hline
	{SSL certificate distribution} & 15 \\
\hline
\end{tabular}
%\vspace{-4mm}
\label{table:ssl-lat}
\end{table}

\subsection{Client side impact}

% Firefox: 113.0.2 (64-Bit)
% OS: macOS Ventura 13.3.1 (a)
% RAM: 16GB
% CPU: Apple M2
% Netzwerk: (es tut mir leid ;) ) WLAN über FAU.fm mit 802.11ac 5GHz
% Alle Messungen sind nach dem Prinzip gemacht, das wir besprochen haben also Seite aufrufen, Messen, Cache Löschen und von vorne.
%
% Das heißt, der "plain" Durchschnitt beträgt jetzt auch 32ms, ich weiß nicht, warum der vorhin bei 170ms lag. Vielleicht internes Firefox caching, dass nichts mit Webseiten zu tun hat / warmlaufen?

% unknown:
% loadReport: 26.5 | [25, 24, 27, 27, 28, 27, 25, 27, 26, 27, 27, 26, 24, 25, 28, 30, 27, 26, 29, 25]
% loadVCEK: 987.4 | [767, 781, 440, 720, 774, 759, 748, 723, 978, 722, 417, 6013, 803, 791, 414, 725, 459, 1121, 747, 846]
% validate: 14.75 | [15, 15, 17, 15, 17, 15, 16, 15, 16, 15, 15, 16, 15, 15, 16, 15, 16, 17, 1, 13]
% total: 1076.05 | [860, 868, 523, 810, 855, 844, 836, 815, 1064, 819, 502, 6105, 899, 882, 500, 818, 549, 1212, 829, 931]
%
% known:
% known: 95.95 | [78, 231, 231, 75, 81, 95, 81, 80, 82, 80, 85, 78, 75, 85, 73, 77, 85, 88, 79, 80]
%
% plain:
% plain: 32.6 | [23, 25, 32, 31, 25, 23, 24, 33, 122, 36, 31, 29, 26, 26, 35, 23, 26, 23, 25, 34]

To determine the impact of performing remote attestation and constantly securing the connection of a web browser we evaluate a mobile user scenario. 
The client-side consisted of a notebook (Apple M2, 16 GB RAM, macOS Ventura 13.3.1) connected via wireless to the \magic-protected Boundary Node. 
An instance of Firefox (13.0.2) equipped with and without our web extension repeatedly accessed a minimal web page (see Table~\ref{table:webextension}) using Selenium (4.12). 
The base network latency accounted for 5.2\,ms and the plain access of the web page with  100.9\,ms.
In a fresh web session, remotely attesting a \magic \ac{VM} takes (including accessing the web page) 778.9\,ms on average. 
Thereby contacting the AMD key server for the VCEK consumes most of the time (427.3 ms).
Since the VCEK is the same until the \ac{SNP} firmware is updated, it can be cached, and this speeds up the access of websites that are frequently visited. 
Once the remote attestation succeeds, it has to be monitored that the connection is not reset and replaced with a new certificate. 
This requires for each request to query the browser from the web extension for the connection context. 
Accessing the test web page while a remote attestation has already been performed requires on average 115.0\,ms. 
If the browser itself would be modified this overhead could be eliminated, as a re-establishment of a connection could simply trigger a re-validation in this situation. 

\begin{table}[t]
\caption{Browser-based remote attestation and validation }
\centering
\begin{tabular}{|l|r|} 
 
 \hline
 & \textbf{Latency (ms)} \\
 \hline
 \hline
	{Network latency} & 5.2  \\
\hline
   {Plain HTTP GET} & 100.9 \\
\hline
	{HTTP GET and remote attestation } & 778.9\\
\hline
	{HTTP GET and conn. validation } & 115.0 \\
\hline
\end{tabular}
%\vspace{-4mm}
\label{table:webextension}
\end{table}

%!TEX root = ./main.tex

\section{Related Work} 

%Remote attestation is essential for securing sensitive workloads via trusted execution.
%Often there is one remote party that validates a secured execution context and, if successful, further takes possession. 
%This can be achieved by injecting decryption keys for configuration and service data stored on an encrypted volume. 
%End user of such TEE-secure workload can at best indirectly deduce that trusted execution is used by the fact that a secure connection can be established indicating that the service owner has successfully performed remote attestation and injected a trustworthy certificate.  
%In such a setting the end user needs to fully trust the service provider, \magic targets to reduce this control by exposing remote attestation to the user and structuring the execution context appropriately. 
There has been extensive research regarding remote attestation of TEEs~\cite{opera, fl-ra, ws-at, hydra, watz, sev-sgx-at, cas, uni-att} and how it can be used to establish trust in software components and security architectures.
%\magic targets VM-based TEEs and how to enable end-users to receive evidence about the confidential services they are using without depending on the cloud or service providers.
%
Ryoan~\cite{ryoan16osdi} and TrustJS~\cite{trustjs17} feature a two-way sandbox, protected inside SGX enclaves, enabling end-users to run their workloads on untrusted systems and attest the execution environment.
%The runtime environment can load confidential code only known to the service provider and cannot be accessed or manipulated from outside the \ac{TEE}. 
%The confidential code can securely receive user data and process it. 
Due to the construction of the double-sided sandbox, the execution results are only exposed to the end-user.
So the data of users stay confidential to the service provider. 
\magic presents a more flexible approach for VM-based TEEs giving the means to end-users to validate a utilised service including its configuration to the extent of interest without depending on the cloud or service providers. 
%
%Troxy~\cite{bli2018troxy} implements a secure proxy via trusted execution to hide the complexity of BFT fault tolerance from end-users and their legacy applications. 
%Therefore, Troxy shares the idea of protocol translation with one of the example applications, the boundary nodes of the IC. 
%However, as in a commodity setting the owner of the replicated system has full control over the \ac{TEE}. 
%
%Web-based trusted execution as proposed by TurstJS~\cite{trustjs17} features a similar idea to Ryoan by implementing a double-sided sandbox.
%JavaScript code can be executed integrity protected and confidential to the end-user and only access data within the scope of the loaded web application. In this case, the web application provider performs remote attestation on a standardized environment that implements a JavaScript execution environment. 

Remote attestation via TLS~\cite{knauth2018tls} and RATLS~\cite{ratls2022walther} aim to seamlessly combine remote attestation with the TLS protocol. Both approaches could be integrated with \magic.
Narayanan et al.~\cite{snp-tpm} showcase an implementation of a vTPM based on which they perform remote attestation of \ac{SNP} VMs. 
This approach could also be applied to \magic's architecture and enable us to have a runtime monitoring system leveraging the vTPM.
Johnson et al.~\cite{parma} guarantee the confidential execution of containerized workloads by introducing and attesting execution policies which are essentially proof over all the future states of the containers, 
while Pontes et al.~\cite{spire} implement a SPIRE plugin to provision verifiable identities to \ac{SNP} VMs. 
\magic, on the other hand, enables end-users to seamlessly attest a web-facing service, before passing any sensitive data to it, via a web extension that verifies its cryptographic measurement over its loading-time state.

\section{Conclusion}
In this work, we introduced \magic, a novel security architecture that enables end-users to attest web-based services that run on hardware-protected \ac{VM}-based \acp{TEE} from their browsers and to shield their sensitive data from a malevolent cloud or service provider. 
With a small performance cost, our system can be applied in a wide range of use case scenarios, enhancing the security guarantees offered by confidential computing and allowing end-users to reap the benefits.

\begin{acks}
We thank the anonymous reviewers and our shepherd Redha Gouicem for their valuable feedback and Ines Messadi for providing comments on a draft of the paper.
This work was supported in part by Deutsche Forschungsgemeinschaft (DFG, German Research Foundation) as part of the Cluster of Excellence Centre for Tactile Internet with Human-in-the-Loop (CeTI) -- Project ID 390696704, the CRC/Transregio 96 Thermo-energetic design of machine tools -- Project ID 174223256, and the CRC/Transregio 248 Foundations of Perspicuous Software Systems (CPEC) -- Project ID 389792660. 
The authors also acknowledge the financial support by the Federal Ministry of Education and Research of Germany in the programm of "Souverän. Digital. Vernetzt.", the joint project 6G-life -- Project ID 16KISK001K, and the European Union Horizon Europe research and innovation programm under grant agreements 101016577 (AI-SPRINT), 101092644 (NEARDATA) and 101092646 (CLOUDSKIN).
\end{acks}
\bibliographystyle{ACM-Reference-Format}
\bibliography{main.bib}
\end{document}